# Modelling the average income dependence on work experience in the USA from 1967 to 2002

Ivan O. Kitov

## Abstract

The average and median income dependence on work experience and time is analyzed and modeled for the USA. The original data set providing the mean and median income estimates in 10 year long intervals spans a long time period of almost 35 years – from 1967 to 2003.

A microeconomic model linking personal income, population age structure and GDP per capita is used to predict the mean income values in various age groups and their relative evolution in time. Also modeled is the value of work experience where the mean income growth ends and it starts to drop exponentially with increasing age. This work experience increases through time as the square root of the per capita GDP growth.

Prediction for the following 20 years is given for each age group considering potential per capita growth rate of 1.6%. The USA mean income dependence on work experience for 1987 coincides with that for 2002 in the UK – the years when per capita GDP were equal in the countries.

## Introduction

This is the fourth paper from a series presenting results of micro- and macroeconomic modelling of the personal income distribution (PID) in the USA. The first paper (Kitov 2005a) presents the models and describes defining parameters corresponding to the best fit between observations and predictions. The second paper (Kitov 2005b) discusses some principal results of overall PID modelling. The third paper (Kitov 2005c) presents age-dependent PIDs' modelling. The second and the third papers use the same data set provided by the US Census Bureau (2004a). This is a high-resolution data set presenting the personal income data in two time scales: calendar years between 1994 and 2002 and work experience from 0 years to 60 years.

The fourth paper analyses and models a different set of data – the average income dependence on work experience available from the same US Census Bureau web-site (2004b). This data set spans a longer time period of almost 35 years – from 1967 to 2001. Due to some major changes in the income survey procedure consistently applied before and after 1967, as described in the US Census Bureau technical documentation (2002), accuracy of the measurements was progressively improved. No relevant data before 1967 are published, however.

## 1. Mean income as a function of work experience

The mean personal income distribution in 10 year age intervals is available from 1967 to 2001 (US Census Bureau 2004b). Corresponding table contains mean personal incomes for male and female separately. Additional efforts are necessary for obtaining gender independent estimates. Fortunately, the averaging bases, i.e. the numbers of people with income, are also listed for each gender and each age group. For the youngest age group, the data are available only from 1974, and for the oldest age group, from 65 to 74, only from 1987. This is induced by major changes in the Current Population Survey (CPS) procedures. The table presents both current dollars and chained 2001 dollars estimates. Figure 1 illustrates the PIDs in current dollars. A semi-logarithmic scale uncovers two principal characteristics of the distributions: an exponential character for the mean income growth and the roll-off with work experience, and the existence of some critical work experience, $T_{cr}$. This critical time point divides the distribution into two branches: a part increasing as a function like $(1-exp(-\alpha t))$, and an exponentially decreasing branch beyond the time of critical work experience.

Figure 2 shows the same income distributions in chained 2001 dollars. The only observed difference from the curves in Figure 1 is in the distribution levels. Real income changes are less intensive than nominal; the mean real income in the age group from 45 to 54 years has increased only by 1.5 times from 1967 to 2001. Also, there is a smaller real income change observed between 1967 and 1991 compared to the change between 1991 and 2001, despite the real GDP between 1967 and 1991 having increased by 2 times, and between 1991 and 2001 by only 1.4 times. This discrepancy is associated with a change in the relative number of people with income during these years. Figure 3 presents a ratio of the number of people with income (as defined by the Census Bureau 2004b) in the studied age groups and the total number of people (US Census Bureau 2004c) in the same age groups. There was a strong increase of participation factor (relative number of people with income) from 0.8 to almost 1.0 from 1967 to 1980 in all the age groups except the youngest one. The latter group has an almost constant participation factor of around 0.75 throughout. When corrected for the participation factor the mean income distribution looks more consistent with the observed monotonic growth of real GDP (Figure 4), i.e. no convergence of the curves is observed. Here one can assume a same deficiency in the current income estimation procedure as discussed in the previous papers of the series. It estimates only the external sources of income and does not inquire into the household internal distribution of income – a major income source for poor and young people.

A substantial increase of the workforce (the number of people with income) between 1975 and 1981 did not cause a proportional real GDP growth during these years. The effect of the increase looked like the same total income was distributed between a larger number of people: the average income in all of the age groups changed proportionally to the real GDP. In other words, the internal or implicit distribution of income became the external or explicit one, but the production of the total income does not depend on an implicit or explicit mechanism for the income distribution. Thus, Figure 4 presents a natural mean income distribution (total income in a given age group divided by the total population in this age group), while Figures 1 and 2 show biased distributions: the baseline number of people is arbitrarily defined by the income survey procedure or changes with the time of the survey questionnaire. When a new and different definition of income is applied, as it happens with a progressive improvement in survey procedure, this baseline number undergoes some changes and one can not easily compare results of two consecutive surveys. The natural mean income distribution is based on a definition of the personal income which is consistent over time, however. The personal income is the amount of money (in various forms), which a person can spend by her/his own decision. This balances income and expenditures in every age group and overall. The definition effectively considers all the population of a given age.

The largest correction has to be applied to the youngest age group. This group is also the most affected by the income estimate procedure because a larger part of young persons incomes comes from internal sources. One can expect a higher average income when these sources are included. The income from these sources, obviously, allows the people without income to survive.

The internal sources are very important for some age groups and individuals, but do not substantially change the overall distribution. The income obtained from the external sources and estimated during the CPS is much larger than the redistributed one. Thus, one can expect some bias in the youngest age group and a more accurate representation in other age groups and on the whole.

The corrected personal income distribution is taken to constrain the macroeconomic model. This distribution includes all people of 15 years of age and over and avoids some deficiencies in the CPS estimates based on a limited consideration of the personal income. On the other hand, the distribution corrected for people without income results from direct measurements of income and population, i.e. represents a valid object for a numerical study and modelling. The distribution has the same features as the original current dollar and chained 2001 dollar distributions: exponential growth and fall, and the presence of a critical age, where growth turns to decay. The latter parameter plays a key role in understanding of the evolution of personal income distribution.

Figure 5 displays the mean income distributions normalized to the peak mean income value of the corresponding year. Thus, the peak value of the normalized distributions is always equal to 1.0. The shape of the normalized distributions evolves in time with a tendency of the critical age, $T_{cr}$, to grow. Figure 6 presents the normalized mean incomes in the studied age groups as a function of calendar year. The peak mean income belongs to two age groups: from 20 to 29 and from 30 to 39 years of work experience. The latter took the lead in the middle 1980s. Previously, the highest mean income was observed in the younger of the two groups. During the late 1960s, the older group also had some years of superiority. When interpreting the difference between the two groups, however, one has to take into account the decreased accuracy of the past estimates and substantial changes in the estimation procedure during last 40 years. The last such change was implemented in 1994. Since then all the curves grew with no reasonable explanation of the phenomena.

One can estimate where the peak mean income value was or will be in various age groups with time. Figures 7 through 10 present results of linear regression analysis of the available normalized mean incomes. The obtained linear dependencies are extrapolated to intercept a unit line, i.e. to reach a peak value among the age groups. In the youngest age group, the coefficient of linear regression is -0.004. This coefficient gives the estimated time of interception around 1790. When a regression coefficient of -0.075, as obtained in other age groups, is used, the intercept time moves to the beginning of the 20$^{th}$ century. Thus, the youngest people had a dominating income position (in terms of mean income) in the 19$^{th}$ century.

In the age group from 10 to 19 years of work experience, the linear regression coefficient is -0.075. Hence, the group was at the top of the income pyramid until the late 1940s. The group from 20 to 29 years of work experience had the peak mean income value until the middle 1980s. This was the only change of the age group with the peak mean income value measured by the US Census Bureau. During the last 20 years the peak mean income has had a tendency to move towards the group with 40 to 49 years of work experience. One can expect the group will take the lead in 2015.

The critical work experience value defines the average rate of economic development. During the last 55 years the real GDP growth trend or economic potential was exactly equal to a reciprocal of $T_{cr}$ (Kitov, 2006) In fact, if the mean personal income grows during 50 years from zero to its peak value, one can suppose that an annual growth would be 1/50 or 2%. The current $T_{cr}$ value in the USA is about 40 years. Thus, the current economic trend in per capita real GDP is 2.5%. During the 1950s, when $T_{cr}$ was about 25 years, the economic trend was of 4%.

Because the averaging time intervals are relatively wide (10 years), it is hard to determine an exact critical work experience value from a single distribution. The set of distributions, however, reveals changes in the critical work experience value. It was assumed (Kitov 2005a) that the critical value is proportional to the square root of the per capita real GDP growth:

$$Tcr(i)=Tcr(i-1)*sqrt(1+dGDP) \quad \textbf{(1)}$$

where $T_{cr}(i)$ is the critical work experience value for the year $i$, $T_{cr}(i-1)$ is the same for the previous year, and $dGDP=(GDP(i)-GDP(i-1))/GDP(i-1)$ is the real per capita GDP growth rate. As before, the real GDP growth rate is corrected for the population 15 years of age and over.

Figure 11 demonstrates the evolution of $T_{cr}$ as a function of per capita real GDP according to (1) for the years 1950 to 2002. From this one can compare the predicted and observed values: $T_{cr}$ was about 25 in the late 1950s, reached the 30 year boundary in the late 1970s, and is currently near the 40 years threshold. One has to take into consideration difference between the theoretical $T_{cr}$ given for every single year of age and $T_{cr}$ obtained as a result of averaging in 10 years wide interval. The latter has a lag of several years relative to the former. Unfortunately, there are no direct measurements of $T_{cr}$.

## 2. The average income distribution vs. work experience and its evolution with time

The average personal income distribution covers the longest period of time among the income related data sets, which are available at the U.S. Census Bureau web-site. By definition, average income is sensitive to the number of people in every age group and the personal income distributions within the groups. Detailed data for the personal income distribution are available only from 1994, and the mean income data from 1967. There were many changes in the survey procedure during the years between 1967 and 2001 (US Census Bureau, 2005a). These changes definitely result in varying accuracy for the mean income estimates over the years and are responsible for some sudden changes in the distributions in successive years. The very long averaging time windows also cause substantial variations in the estimated indices of decay above the critical time $T_{cr}$. So, it is hard to match all the observations of the mean income in one model. Figure 12, for example, illustrates the difference in exponential decay characteristics obtained from the 10 year wide and 5 year wide averaging intervals for the year of 2001.

The principal goal of the mean income modelling was to obtain an accurate prediction for the years between 1967 and 2001. In order to fit the observed exponential decay beyond $T_{cr}$ for these years, we fixed the relative income at age of 60 years as equal to 0.84 of the peak mean income in a given calendar year. As presented in Figure 10, the relative income in the age group from 55 to 64 years varies from 0.75 to 0.85 times that of the peak value for the years studied. Other parameters in the model are the same as described in previous papers of the series. Exponential decay observed above the critical time, $T_{cr}$, follows the relationship (3) from Kitov (2005a). The second exponent has a modified index $(-\alpha_1(t- T_{cr})/L)$, where $\alpha_1=-ln(0.84)/(60-T_{cr})$. This makes the curves for each and every year pass through the point (60, 0.84).

Figure 13 displays results of the mean income modelling. A factor used to convert the predicted values into the observed values is the same for both years and is equal to 72. This factor plays no role in the modelling itself and is only used to compare the predicted and observed distributions of the mean income. The observed and predicted curves for 1967 and 2001 almost coincide in the work experience range from 0 to 60 years, except for the youngest age group. This means that the model predicts not only the shape of the individual distributions, but also the observed evolution in the mean income distribution over a long period of time on the basis of independently measured external parameters. This uses a very simple relationship between a few internal parameters, which can be accurately calibrated to match various observations.

Figure 14 presents the same curves for the years 1987 and 1974. These years are characterized by a considerable change in the CPS procedure (US Census Bureau 2005a). One can expect that this change in the survey method might lead to some change in the mean income level: when more sources of income are used in averaging a larger mean income is expected. The figure reveals this change – the conversion factor in these two years is 79, i.e. about 10% larger than in 2001. Other years are also characterized by some variations in the conversion factor, but these variations are small compared to the variation observed for the original values of the mean income (not corrected for the population without income). Figures 15 and 16 demonstrate these strong variations. The conversion factor changes in the range from 76 for 2001 to 95 for 1974. The conversion factor is 90 for 1967. Such variations are induced by a considerable change in participation factor, as discussed above.

Figure 17 and 18 illustrate some modelling results for the years with the detailed mean income distributions in the cases of the corrected and original mean income values, respectively. This model uses a decreasing branch calibration that is different from that used in the modelling of 10 year wide intervals: the mean income level is 0.45 of the peak value at the age 80. Overall agreement between the curves is very good, except two clear outliers in 1994: the observed mean income in the work experience groups between 30 and 34 years and between 40 and 44 years. These outliers are not observed in other years under study and may be associated with the change in the survey methodology in 1994. In the youngest age group, the predicted value of the mean income is higher than the observed one. This effect has been already attributed to survey procedure shortcomings. The original mean income values fit the observations in the youngest group much better due to a severe income underestimation in the survey.

Direct modelling of the critical time, $T_{cr}$ is hindered by the absence of relevant measurements. Precise values of $T_{cr}$ usually reside somewhere between the end points of the 10 year wide intervals used in the CPS survey. Resolution of the age-dependent mean income is too low to accurately measure $T_{cr}$. One can model, however, the time history of the mean income in each work experience interval as presented in Figure 6. The evolution depends on the observed per capita real GDP growth rate and $T_{cr}$ as defined by the macroeconomic model (Kitov 2005a). The evolution is not a uniform increase of the mean income in each age group, but results from a complex interaction between people of various ages leading to the redistribution of income towards the age group with the peak mean income.

The mean income values normalized to the peak mean income of the corresponding year give a representation equivalent to that in Figures 13 through 18, but instead reveal a time dependence not on the work experience dependence described above. Both dependencies are simultaneously defined by the best fit model. Figures 19 through 23 consequently present results for the five studied work experience groups: from the youngest group (from 0 to 9 years) to the oldest group (from 40 to 49 years). Two observed curves (corrected for the population without income and original) and one predicted curve are displayed.

In the youngest age group, the corrected observed curve and the predicted mean income curve diverge considerably in line with the above discussion. The original mean income values are much closer to the predicted ones. The same is valid for the other four work experience groups: the original mean income values are better fitted in relative terms. There is a large discrepancy in the absolute level of the corresponding distributions, however, as discussed above.

The above comparison of the predicted and observed (corrected for the population without income) mean incomes shows an important overall agreement of the curves and a considerable divergence during some relatively short time intervals for some age groups. For example, there is an almost 10% deep trough between 1980 and 1990 in the observed mean income curve in the work experience group between 30 and 39 years. The predicted curve does not show such behaviour and retains its value close or equal to 1.0. There are also 2% to 4% amplitude variations in the observed curve from that predicted for the work experience group between 20 and 29 years. This discrepancy can be partly related to the CPS procedures and changes in the population estimates related to decennial censuses. The latter can reach several percent in some age groups. For example, one can compare two different mean income estimates obtained in 2000 and based on two different population estimates (US Census Bureau 2005a).

Thus, there are some concerns about the accuracy of the mean income estimates. In fact, in order to derive the mean income from the survey one has to know the exact PID. Any error in the high-income end of the PID leads to a large error in the mean income because of larger relative input of the high-income population in the net income. The low-income population does not add much to the net income and usually is better presented in the CPS surveys just because it is larger.

Median income may be a potentially more accurate characteristic of income for the various age groups. It does not depend much on the high-income end of the PID – there are not so many people with high incomes to distort the median income. Median income is very close to the low-income end of the PID. All this makes median income a very robust characteristic of the PID and a good parameter to model. As before, the largest modelling problem is observed in the youngest age group where about 25% are excluded from the median income estimate.

The median income estimates are available from the US Census Bureau (2004d). The data set covers the period between 1974 and 2003. An illustration of the difference between the measured mean and median incomes is given in Figure 24, where the overall values are indicated as a function of time. The mean income grows much faster than the median one in accordance with the faster growth of the high-income end of the PID: for the majority it is getting harder and harder to increase income at the same rate with increasing size of earning means as described by Kitov (2005a). Figure 25 illustrates the effect for the two most important groups with work experience: between 20 and 29 years and between 30 and 39 years. These are the groups where the $T_{cr}$ resides during last 50 years. In the work experience group between 20 and 29 years, the ratio of the median and mean income decreases from 0.85 in 1974 to 0.75 in 2002. The ratio will drop further according to the model predictions.

Figures 26 through 28 present results of the median income modelling. The model parameters are the same as before. The observed and predicted median values are in slightly better agreement than those for the mean income corrected for the population without income. One can not correct the median income values for the population without income as easily as the mean income.

There is a major problem associated with the accuracy of the mean and median income estimates. We model the observed values and obtain estimates of the defining parameters, which correspond to the best fit model. Any error in the observed values transforms to an equivalent error in the defining parameters. On the other hand, the parameters obtained in the previous two papers are the same as obtained from the modelling of the mean and median income measured during a relatively long period of time. This validates, to some extent, the model and the observations. Hence, one can conclude that overall the modelling is very successful in spite of some major and minor problems remaining in the observational and modelling parts. The author considers the obtained results are a strong confirmation of the model. Moreover, the model reveals weak points in the current procedure for measuring personal income estimates and provides a good foundation for future corrections and improvements.

The model provides an opportunity to extrapolate the observed behaviour in the future. The observed per capita real GDP growth rate has inertial trend equal to the reciprocal value of $T_{cr}$ less population growth (Kitov 2005d). The trend is about 1.6% during next 20 years. We used this value to predict the evolution of the mean income. Figure 29 presents some predicted curves of the real mean income vs. work experience with five year spacing. Evolution of the population for these years is taken from the US Census Bureau (2005b) population projections. The curves allow an estimation of the total real GDP as an integral over the work experience, .i.e. for the total population 15 years of age and over.

## 3. Conclusions

A comprehensive study of the US personal income distribution and detailed modelling of the principal characteristics of the distributions is carried out and presented in the four papers already mentioned. The principal finding is that people as economic agents producing and earning money are distributed in a fixed and hierarchical manner resulting in a very "rigid" response for the income distributions to external disturbances, including inflation and real economic growth.

The PIDs corrected for the observed nominal per capita GDP growth with time show a very stable shape over the time period from 1994 to 2002. This shape stability is interpreted as an existence of an almost stable income distribution hierarchy in the society, which might be developing very slowly with time. Inflation represents a mechanism compensating disturbance of the PID caused by the real economic growth, i.e. it earns out of the poor people advantages obtained from the real economic growth.

The PIDs in selected age groups, ten and five years long, are characterized by a fast change with age in the beginning and a practical stagnation in shape for older age groups. The youngest age groups reveal an almost exponential PID over the whole income range, while the older groups are characterized by a constant branch and a power law PID roll-off above some threshold value. At higher incomes, the total PID and PIDs for the age groups show a power decrease with an exponent close to -3. The number of people governed by this distribution depends on age. Two branches of the dependence are clearly distinguished – exponential growth and exponential decrease. The number of people with income above $100,000 increases linearly with time in accordance with a power law decay for the PID with an exponent of -3. The mean income dependence on work experience is also characterized by two branches: growth proportional to $(1-exp(-\alpha t))$, where $\alpha$ is a small number close to 0.1 and t is the work experience, and an exponential decay after some critical work experience, $T_{cr}$, which develops in time as the root square of the per capita real GDP.

## 4. Discussion

A microeconomic model of the personal income distribution and evolution with time is developed. The model balances two processes – individual income earning and dissipation of this income. The model accurately predicts the overall PID, its evolution and fine features of the PID in various age and income groups. The results obtained confirm that the observed economic growth is a predetermined process governed only by the single year of age population distribution. The personal income distribution is represented as a sole function of the real GDP per capita and population distribution over age. There are no similar data sets in other countries, which might confirm that notion. The only example is given by the UK Inland Revenue (2004). Mean taxable income is expressed as a function of the working experience for years 1999 through 2002. Figure 29 displays a comparison of the UK taxable mean income estimates and the mean income estimates for the year of 1986 in the USA. The curves are similar and correspond to the same per capita real GDP of $26,500 in the UK (2002) and USA (1986).

This study uses the per capita GDP growth as an external parameter. The personal income is shown to be a predetermined function of this parameter. One can interpret this relationship in the opposite direction, however. The personal incomes as a result of each individual's effort to earn (or produce) money represent the driving force of an economy. It is the sum of the personal income that makes the GDP. So, the per capita real (and nominal) GDP growth rate is unambiguously determined by the current distribution of the personal income, which in turn depends on the population distribution. As obtained above, the mean income distribution (the integral of the mean income over working experience and number of people gives a GDP estimate) is governed only by values at two points - the starting point of the distribution and $T_{cr}$.

Hence, one can suppose that the numerical value of the real GDP growth rate (in developed countries) can be represented as a sum of two terms. The first term is the reciprocal of $T_{cr}$, which is often called the economic trend or potential. The current value of $T_{cr}$ in the USA is 39 years, i.e. the economic trend is 0.027. The second term is inherently related to the number of people of some specific age. In the USA and the UK this age is 9 years. In European countries and Japan it reaches 17 years. This term creates a high frequency variation in the GDP growth rate and is expressed by the following relationship

$$(1/2)(N(i)-N(i-1))/N(i-1)=0.5dN(i)/N(i),$$

where $N(i)$ is the number of the people of the specific age at time $i$, and $N(i-1)$ is the same at a previous time $i-1$. Thus, one can write a relationship for the GDP change:

$$dGDP(i)=0.5(N(i)-N(i-1))/N(i-1)+1/T_{cr}(i-1) \quad \textbf{(2)}$$

where *dGDP(i)* is the real GDP relative growth rate for the given period *dt* (day, month, quarter or year) between times *i* and *i-1*.

Completing the system of equations is the relationship between real GDP per capita growth rate and $T_{cr}$:

$$T_{cr}(i)=T_{cr}(i-1)sqrt(1+ dGDP(i) - dNT/NT) \quad \mathbf{(1')}$$

where *dNT/NT* is the relative change of the total population during the same period of time.

Forward calculations for the USA show excellent results near Census years. This, for example, gives very good predictions of the recession in the last 25 years. There are some large differences, however, between Censuses.

In order to assess the accuracy of the population estimates an inverse calculation was done using the following relationship

$$N(i)=N(i-1)\ (1+2\times(dGDP - 1/T_{cr})) \quad \mathbf{(3)}$$

An example of such an inversion is presented in Figure 30.

Results of the direct real GDP prediction and inverse population calculations for the USA, UK, Germany, France, Japan, Russia, and Austria will be presented in the next paper in preparation.

**Acknowledgements**

The author is grateful to Dr. Wayne Richardson for his constant interest in this study, help and assistance, and fruitful discussions. The manuscript was greatly improved by his critical review.

## Figures

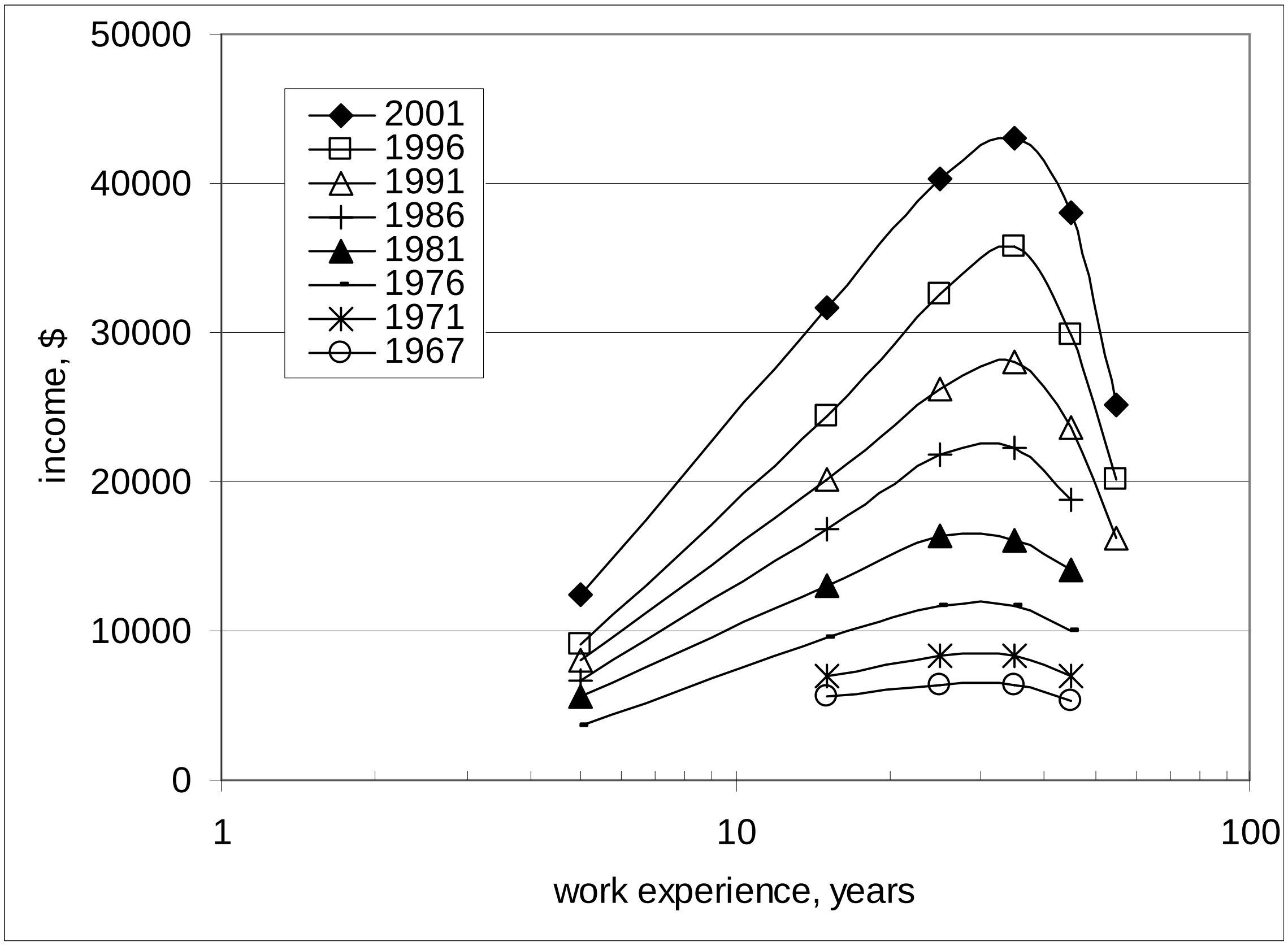

Fig. 1. Mean personal income (current dollars) as a function of work experience for selected years from 1967 to 2001.

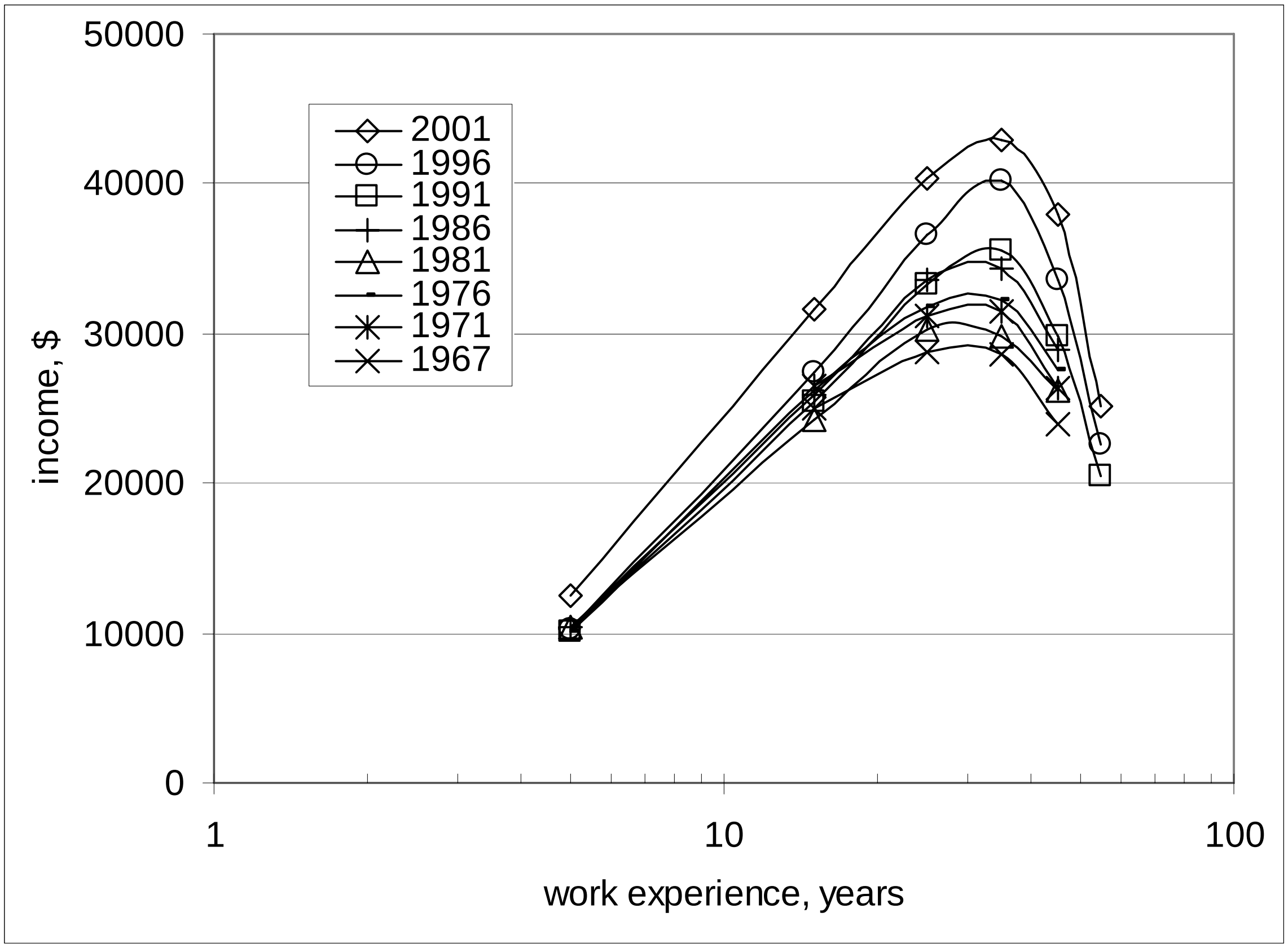

Fig.2. Mean personal income (chained 2001 dollars) as a function of work experience for selected years from 1967 to 2001.

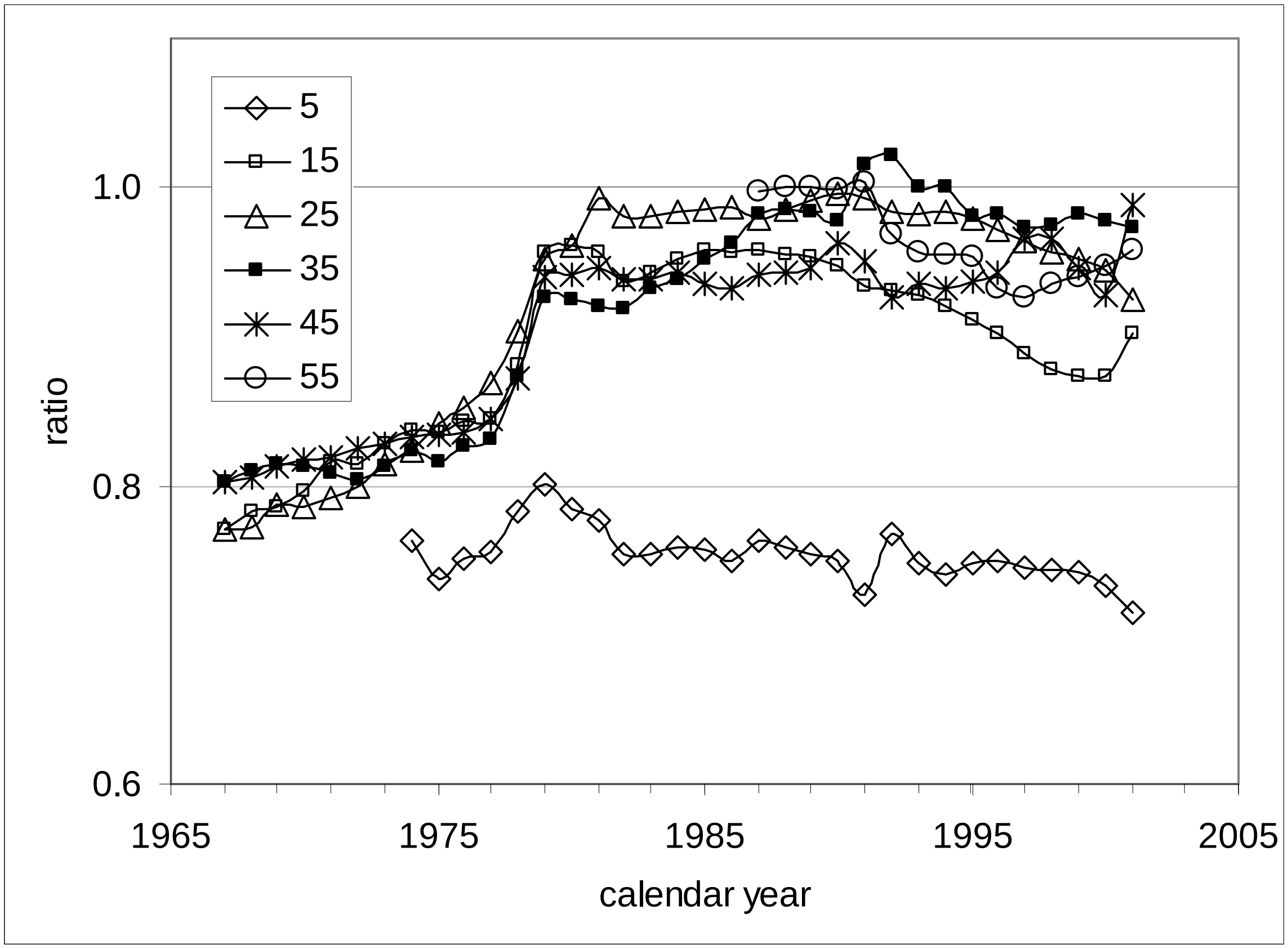

Fig. 3. Ratio of the total number of people and the number of people with income in various age groups. The youngest group is characterized by a participation factor of 0.75. During the late 1970s, participation factor in other age groups increased from 0.82-0.85 to 0.92 -0.99. The ratios are used for the correction of the average personal income in corresponding age groups.

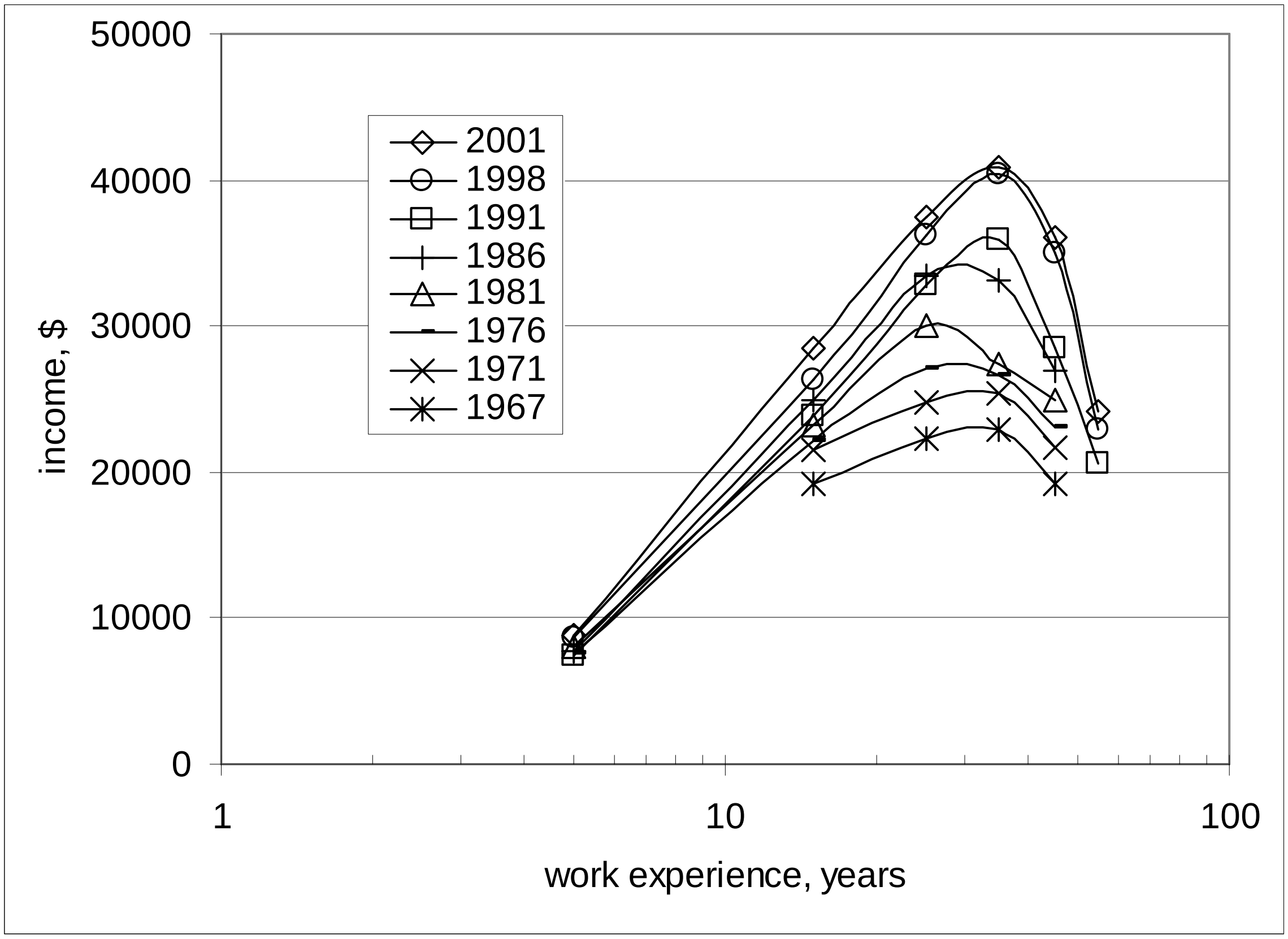

Fig. 4. Mean personal income (chained 2001 dollars) as a function of work experience for years from 1967 to 2001. Mean income is corrected for people with no income.

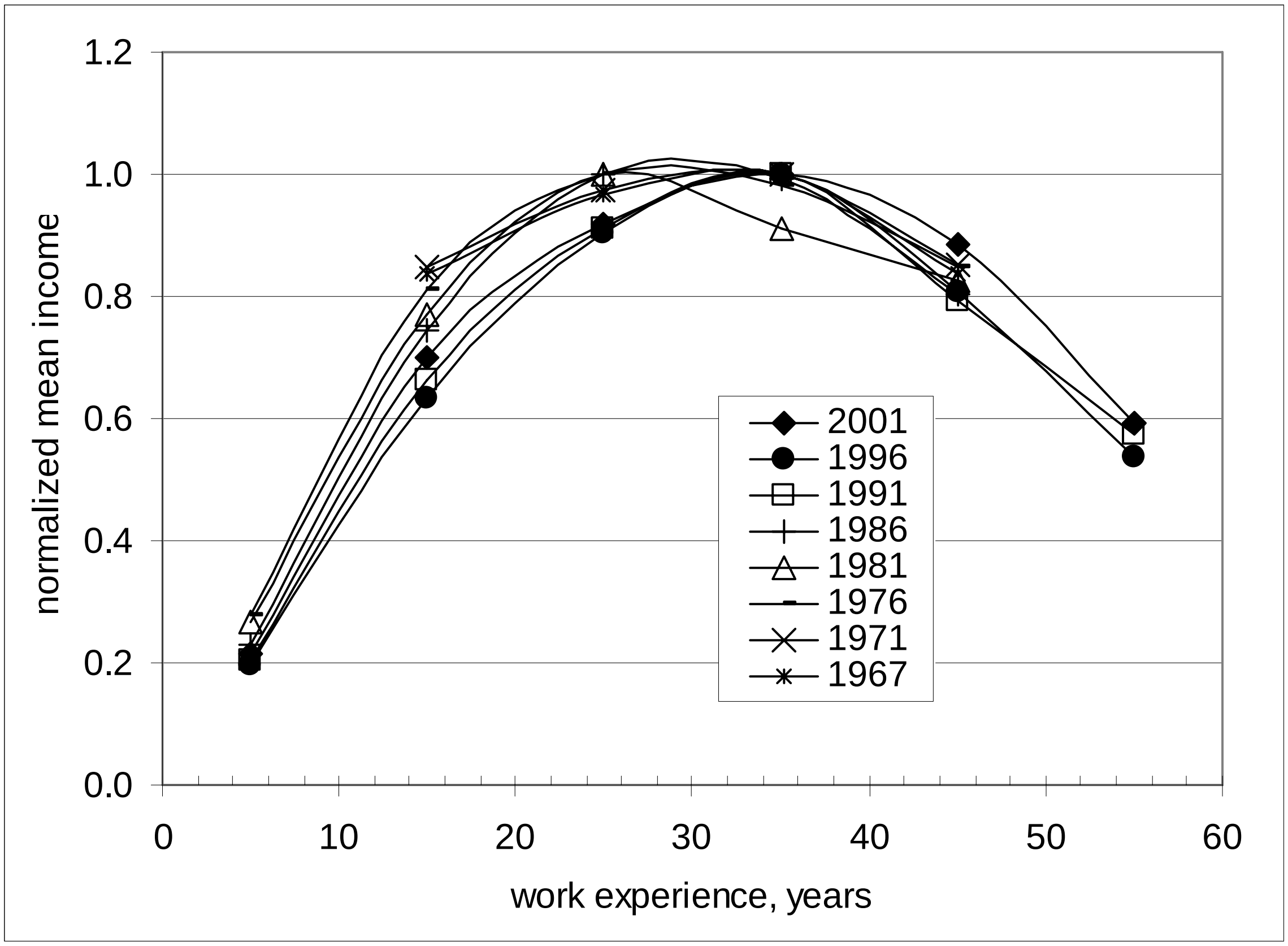

Fig. 5. Normalized mean personal income as a function of work experience for years from 1967 to 2001. The distribution is corrected for people without income.

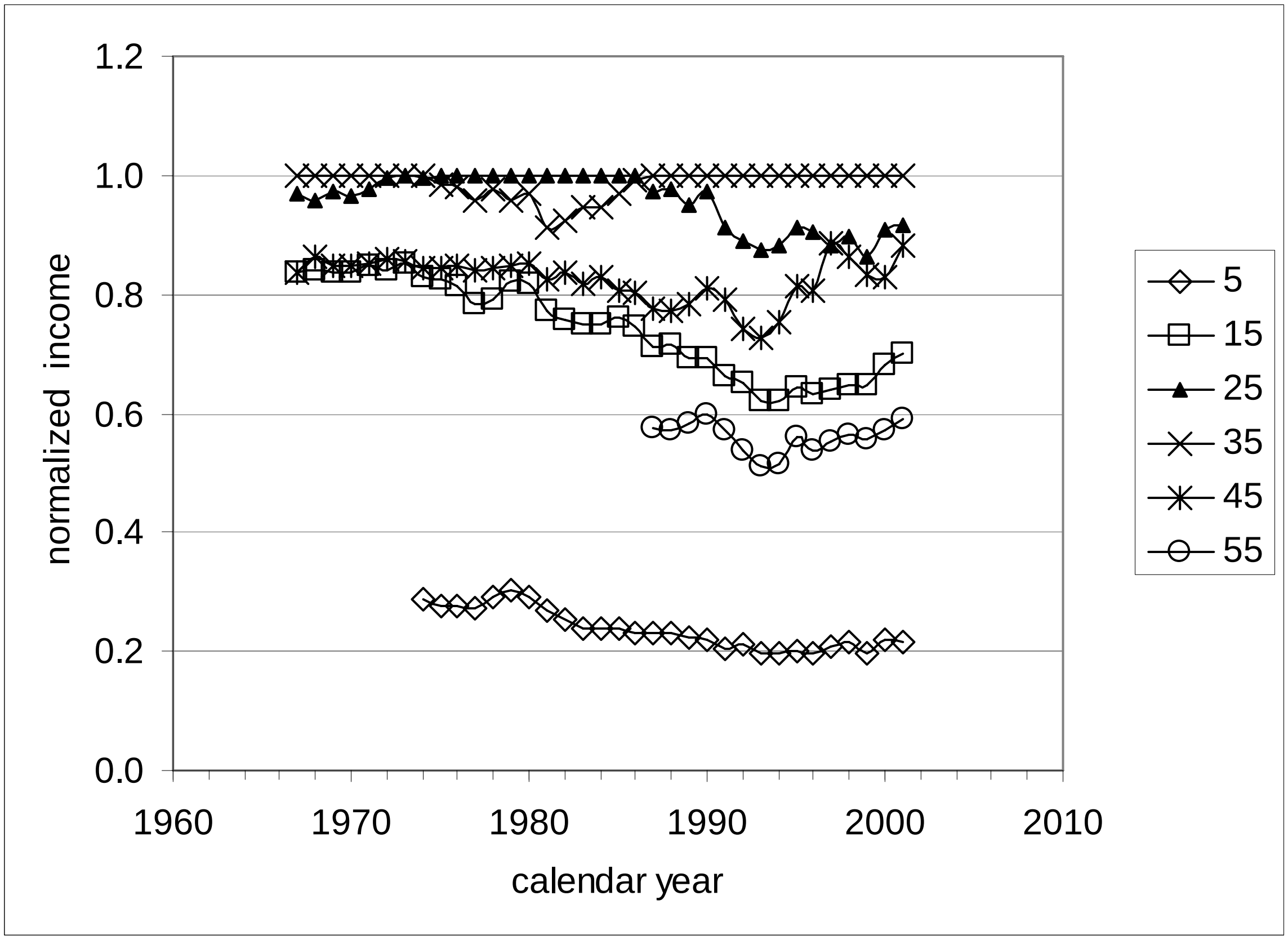

Fig. 6. Mean income in various age groups normalized to the peak value in the corresponding year. The annual peak values are usually in the work experience group from 20 to 29 years (25) or from 30 to 39 years (35). Currently the peak value resides in the latter group.

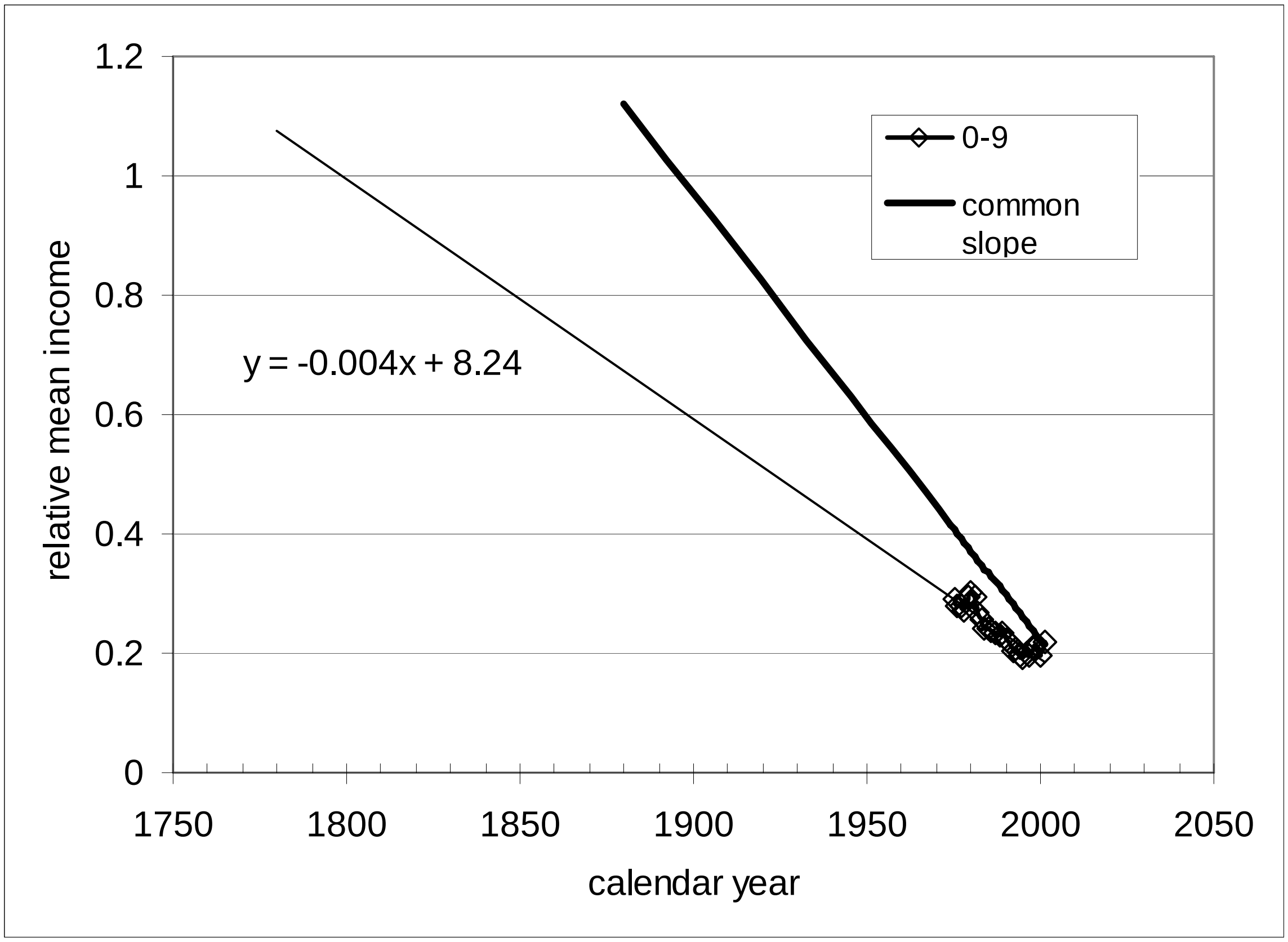

Fig. 7. Mean income in the age group from 15 to 24 years normalized to the peak value in the corresponding year as a function of calendar year from 1967 to 2001. A linear regression indicates that this age group could contain the peak value in the last quarter of $18^{th}$ century. If the slope of -0.075 obtained for the older age groups shown in Fig. 20 and Fig. 21 (common slope) is used, the group could have the peak value near the end of the $19^{th}$ century.

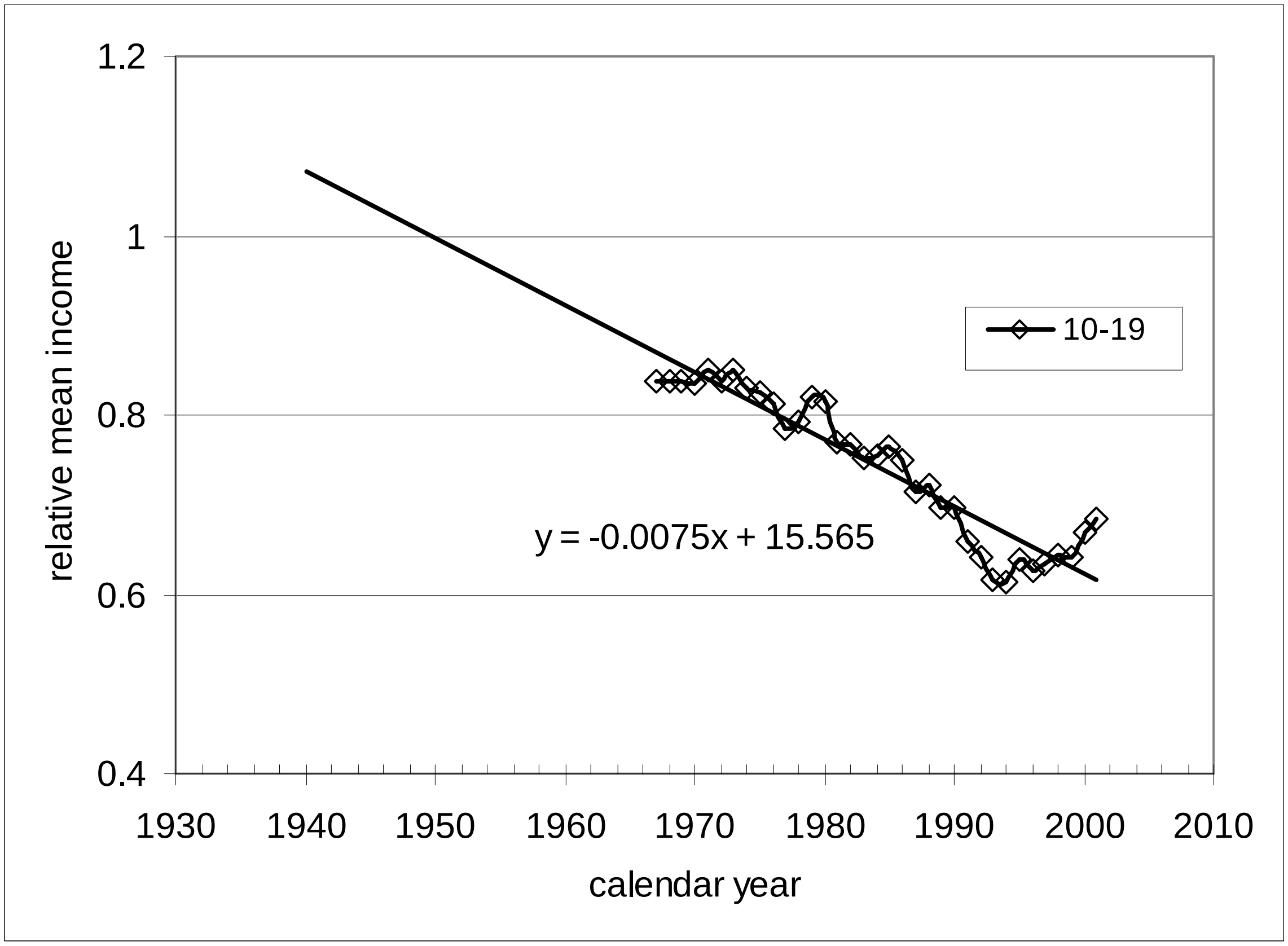

Fig. 8. Mean income in the age group from 25 to 34 years normalized to the peak value in the corresponding year. A linear regression gives a regression coefficient of -0.075 and indicates that this age group reached the peak value between the years 1940 to 1950.

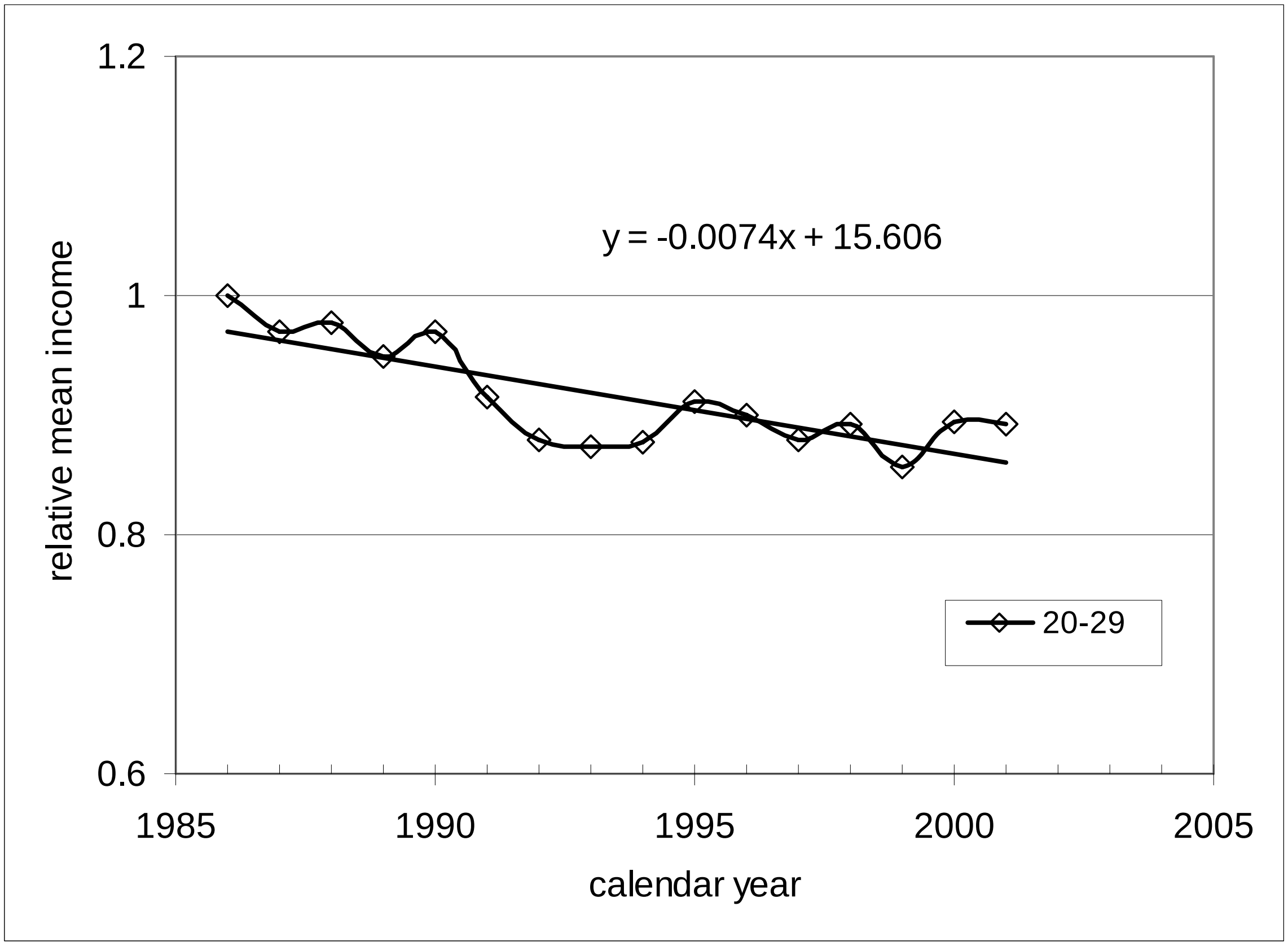

Fig. 9. Mean income in the age group from 35 to 44 years normalized to the peak value in the corresponding year. A linear regression gives a regression coefficient of -0.074. This age group reached the peak value in 1986.

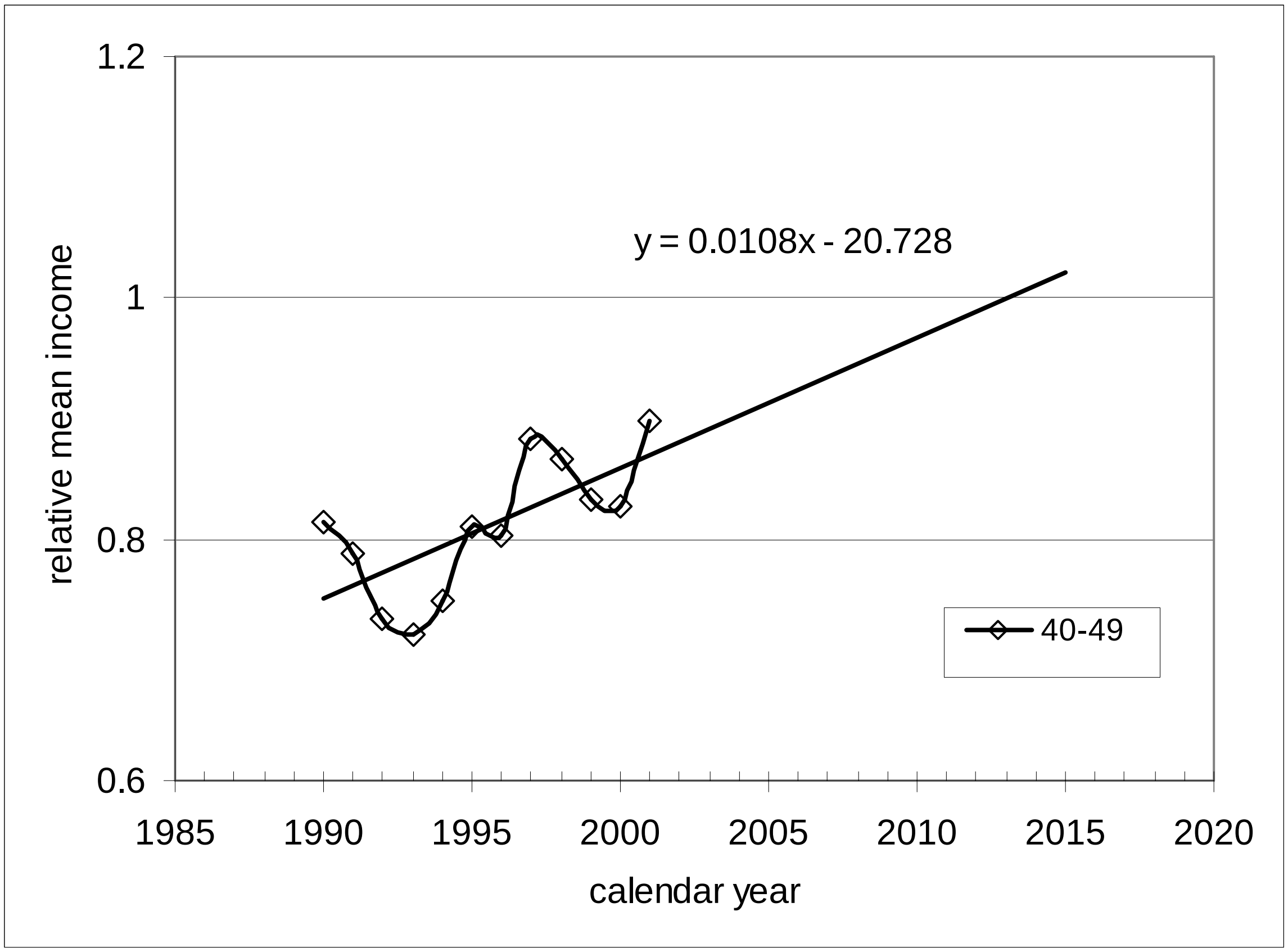

Fig. 10. Mean income in the age group from 55 to 64 years normalized to the peak value in the corresponding year. A linear regression gives a regression coefficient of 0.011. This age group will reach the peak value in the next 10 to 15 years.

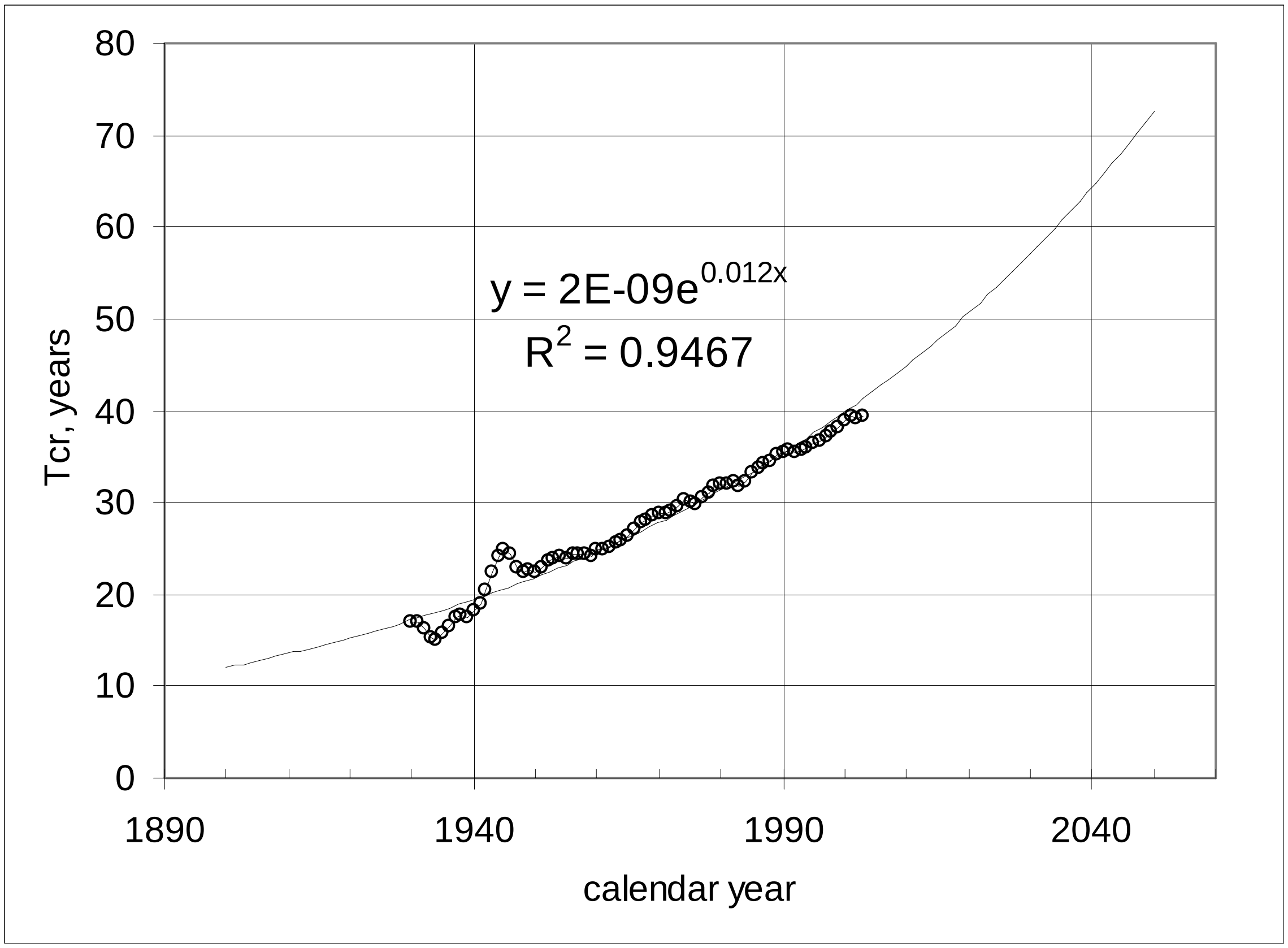

Fig. 11. Critical work experience as a function of calendar years calculated from the corresponding per capita real GDP growth rate using equation (1). $T_{cr}$ was about 25 in the late 1950s, reached the 30 year boundary in the late 1970s, and is currently near the 40 years threshold. One has to take into consideration the difference between theoretical $T_{cr}$ given for every single year of age and $T_{cr}$ obtained as a result of averaging in 10 years wide interval. The latter has a lag of several years relative to the former.

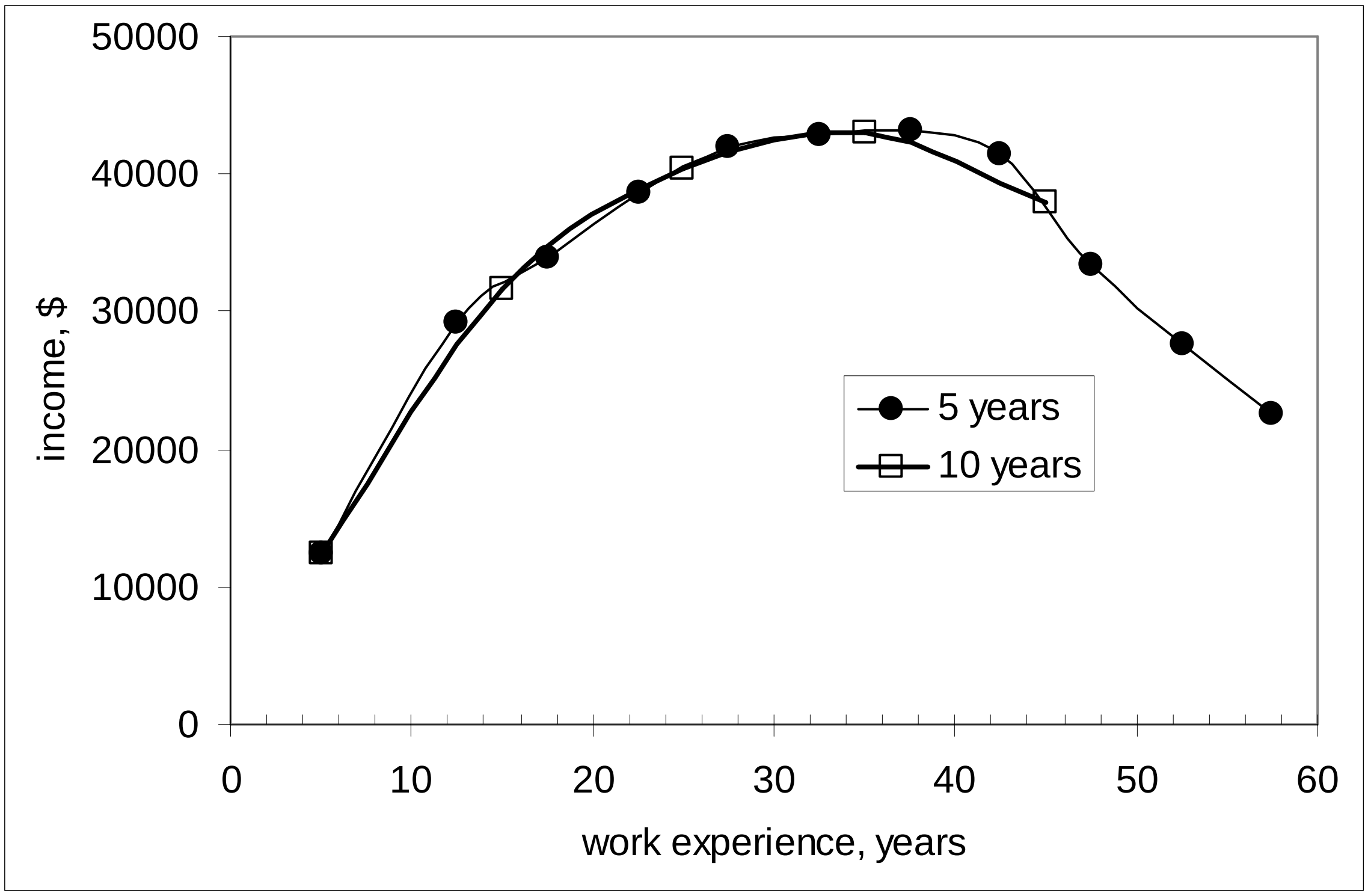

Fig. 12. Comparison of the mean income distributions obtained in 10 year and 5 year wide intervals for the year of 2001. The distributions have quite different asymptotic decay.

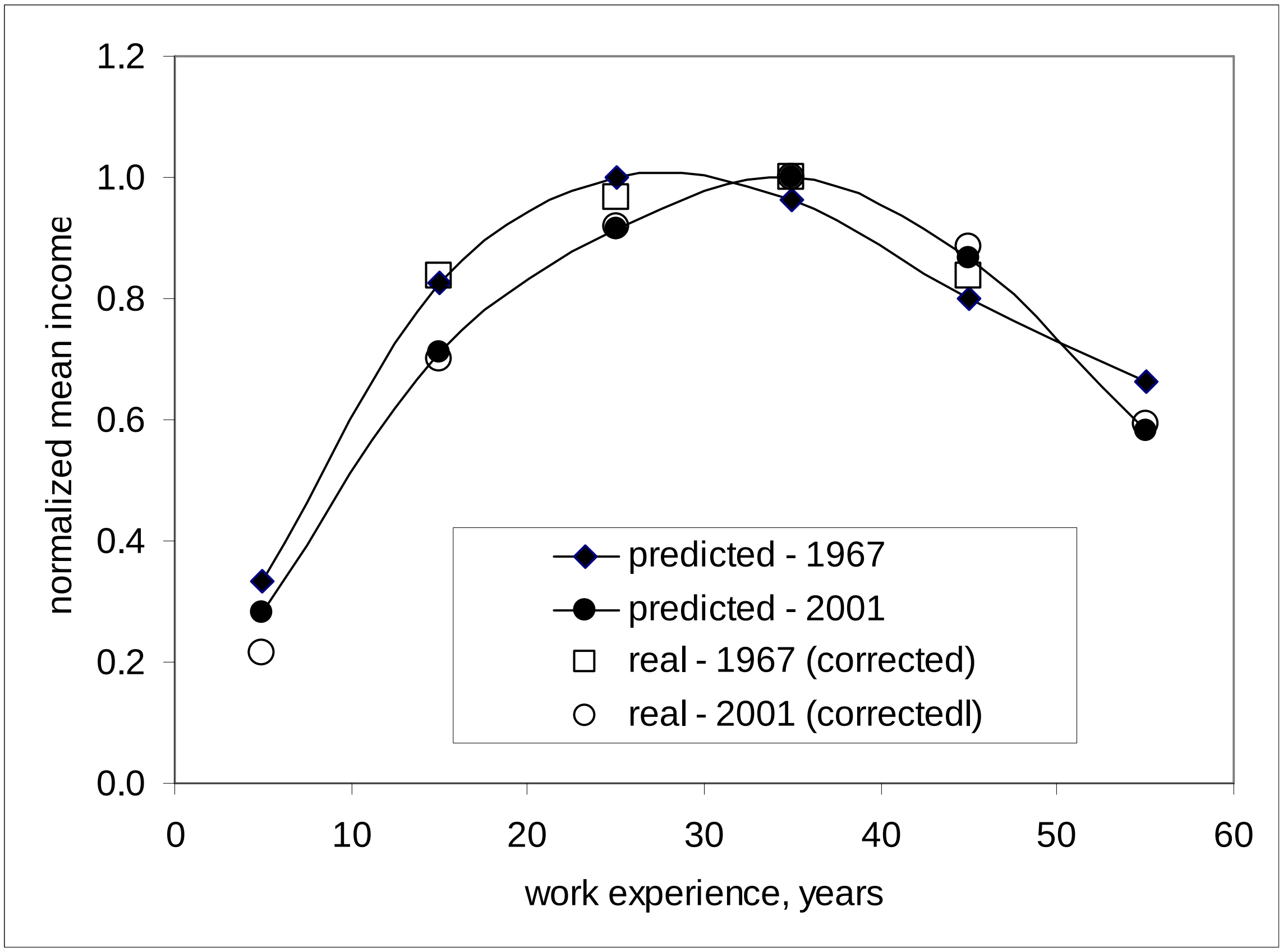

Fig. 13. Comparison of observed and predicted mean personal income distributions for the years 1967 and 2001. The observed mean incomes are corrected for the population without income. Averaging is accomplished in 10 year long intervals of work experience. The boundary condition for the mean income at the age of 60 is 0.84 times the peak mean income value. The factor used to convert the predicted values is the same in 2001 and 1967 -72.

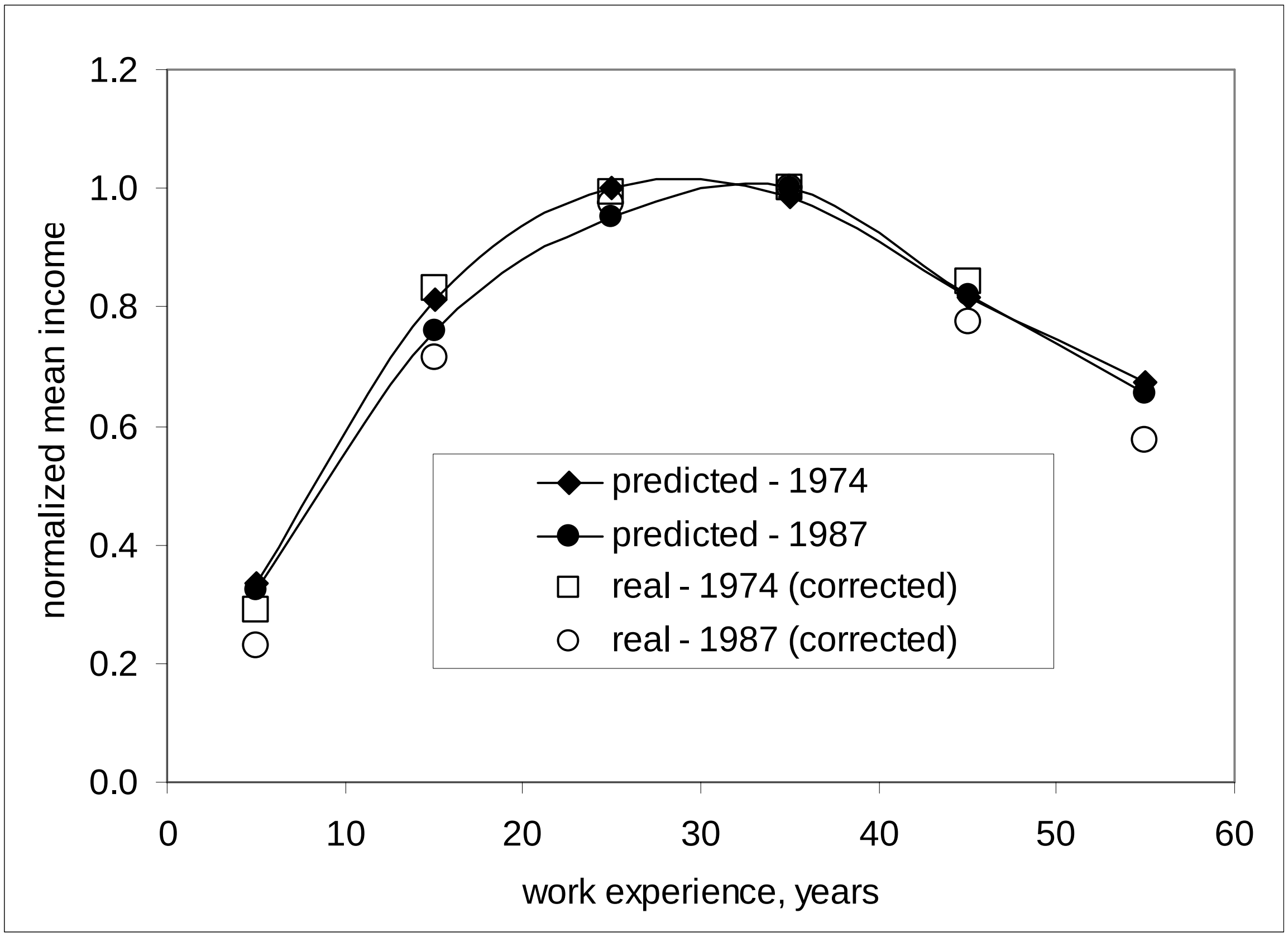

Fig. 14. Comparison of observed and predicted mean personal income distributions for the years 1987 and 1974 (the years of major changes in the CPS procedure). The observed mean incomes are corrected for the population without income. Averaging is accomplished in 10 year long intervals of work experience. The boundary condition for the mean income at age 60 is 0.84 times the peak mean income value. The factor used to covert the predicted values is the same in 1987 and 1974 - 79. This value is different from the value used in 2001 and 1967 (see Fig.13).

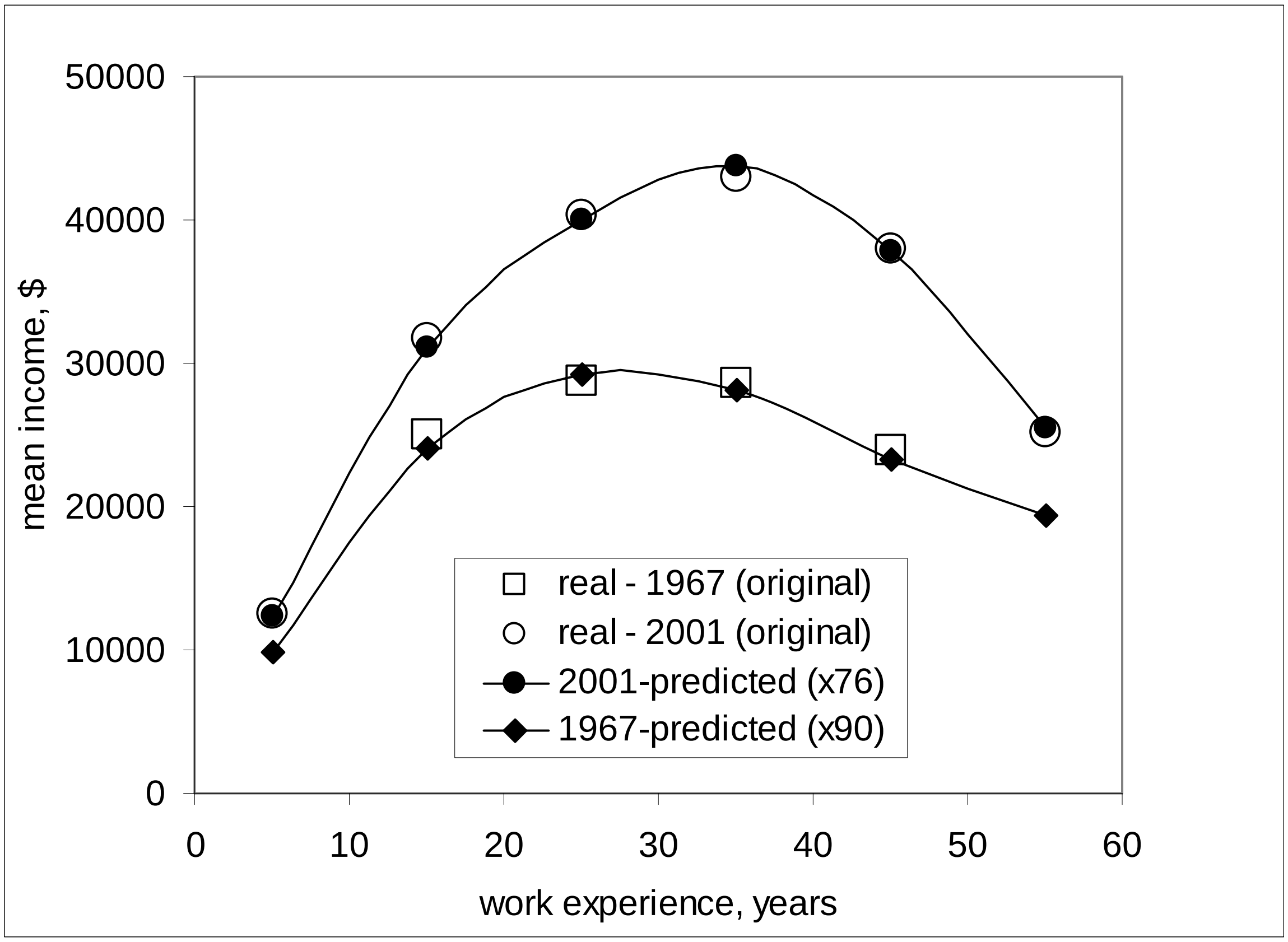

Fig. 15. Comparison of observed and predicted mean personal income distributions for the years 1967 and 2001. The observed mean incomes are as in the original US Census Bureau tables. Averaging is accomplished in 10 year long intervals of work experience. The boundary condition for the mean income at age 60 is 0.84 times the peak mean income value. The factor used to covert the predicted values is 76 in 2001 and 90 in 1967.

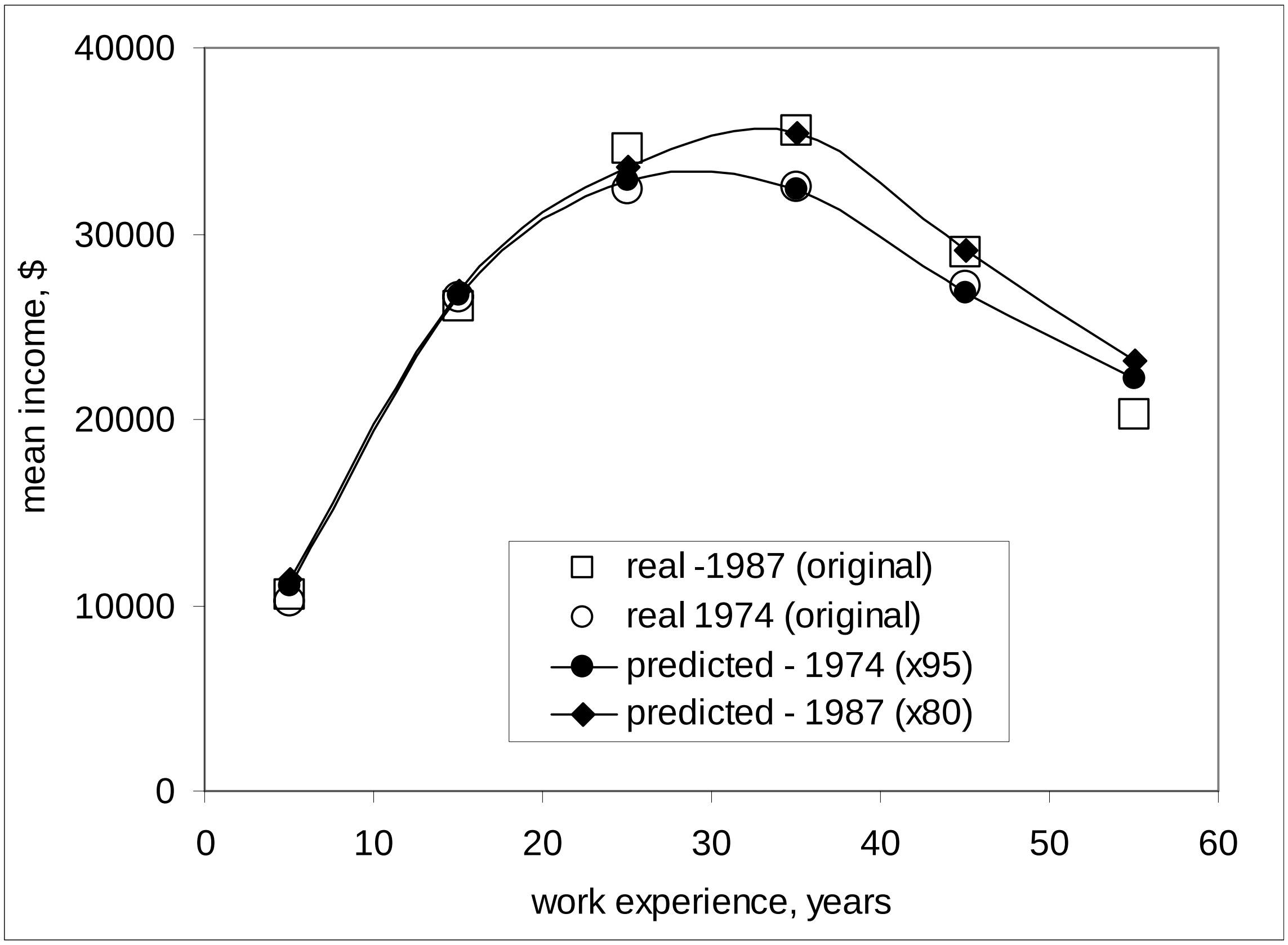

Fig. 16. Comparison of observed and predicted mean personal income distributions for the years 1974 and 1987 (the years of major changes in the CPS procedure). The observed mean incomes are as in the original US Census Bureau tables. Averaging is accomplished in 10 year long intervals of work experience. The boundary condition for the mean income at the age of 60 is 0.84 times the peak mean income value. The factor used to covert the predicted values is 80 in 1987 and 95 in 1974.

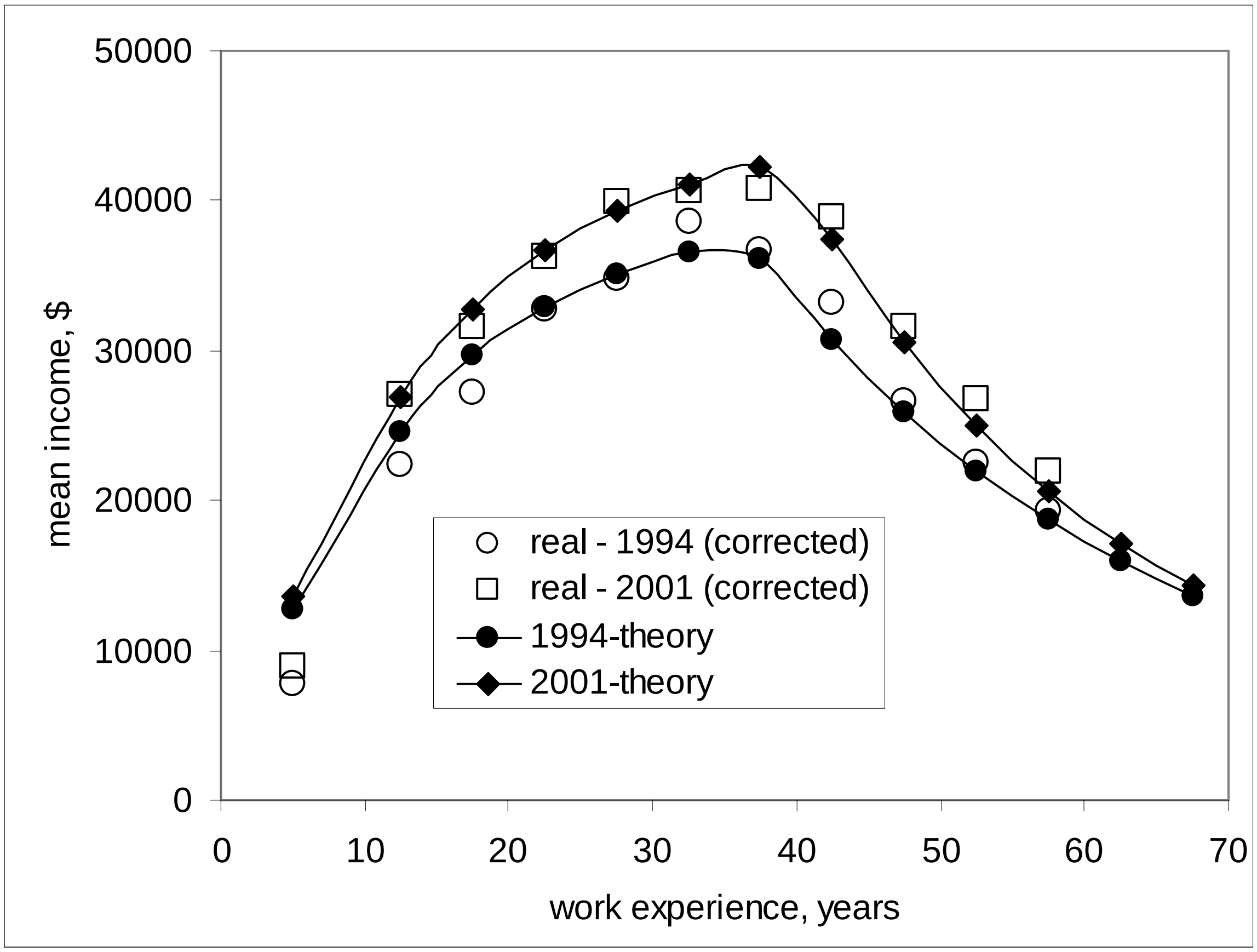

Fig. 17. Comparison of observed and predicted mean personal income distributions for the years 1994 and 2001. The observed mean incomes are corrected for the population without income. Averaging is accomplished in 5 year long intervals (except the first, which is a 10 year long interval) of work experience. The boundary condition for the mean income at age 67 is 0.45 times the peak mean income value.

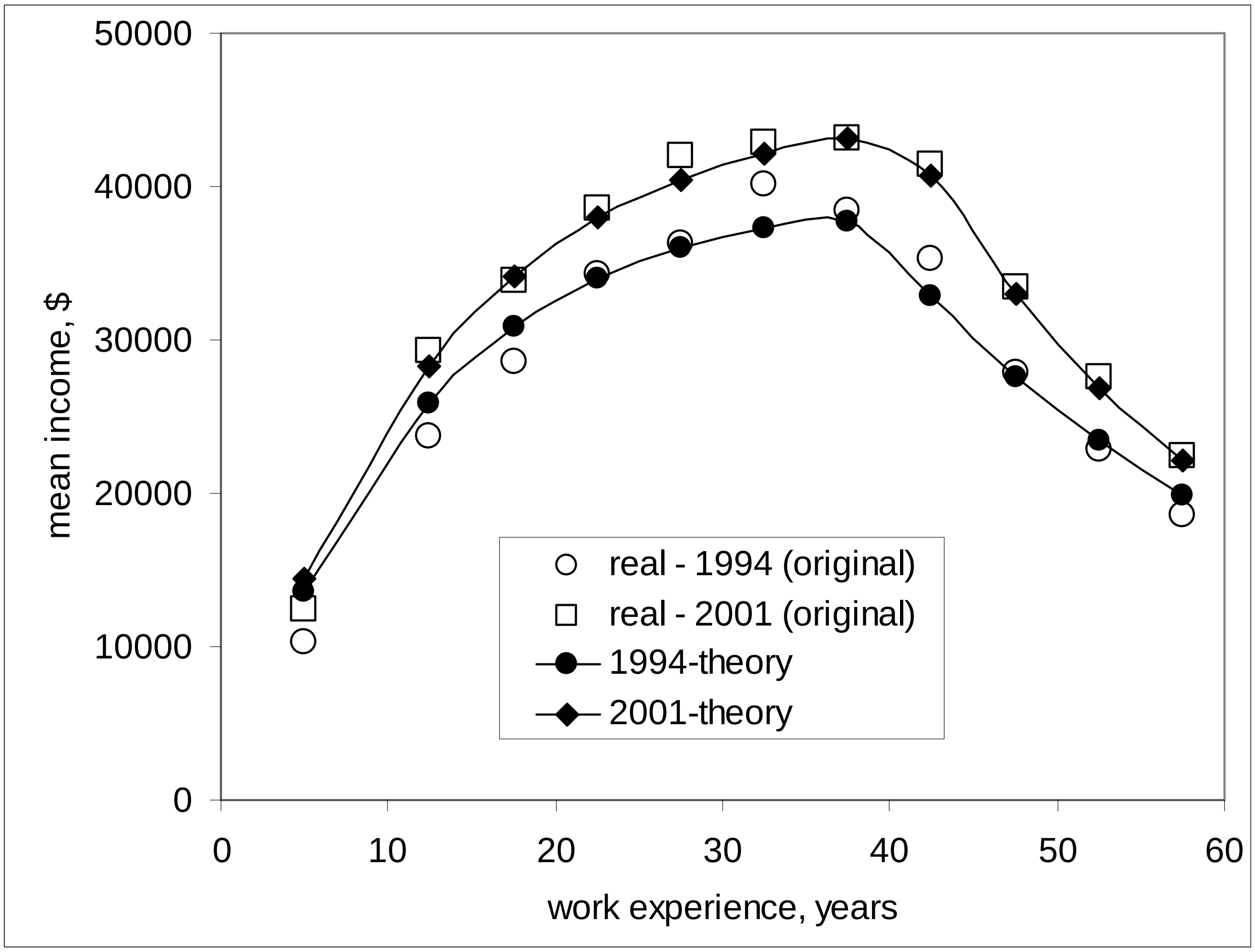

Fig. 18. Comparison of actual and predicted mean personal income distributions for the years 1994 and 2001. The observed mean incomes are as in the original US Census Bureau tables. Averaging is accomplished in 5 year long intervals (except the first, which is a 10 year long interval) of work experience. The boundary condition for the mean income at age 67 is 0.45 times the peak mean income value.

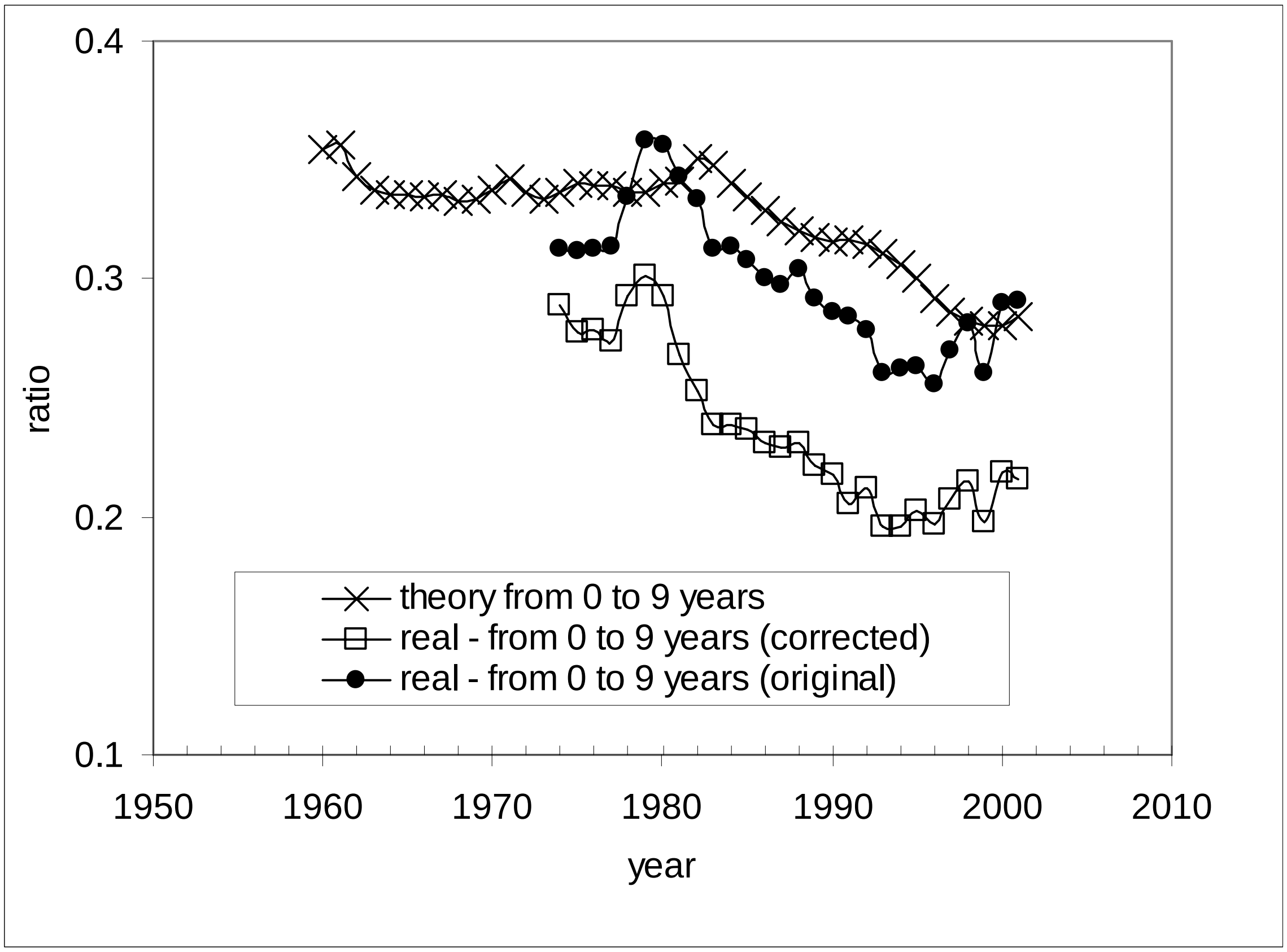

Fig. 19. Evolution of the observed (original and corrected for the population without income) and predicted average income value in the work experience group from 0 to 9 years, normalized to the peak average income over all the work experience groups.

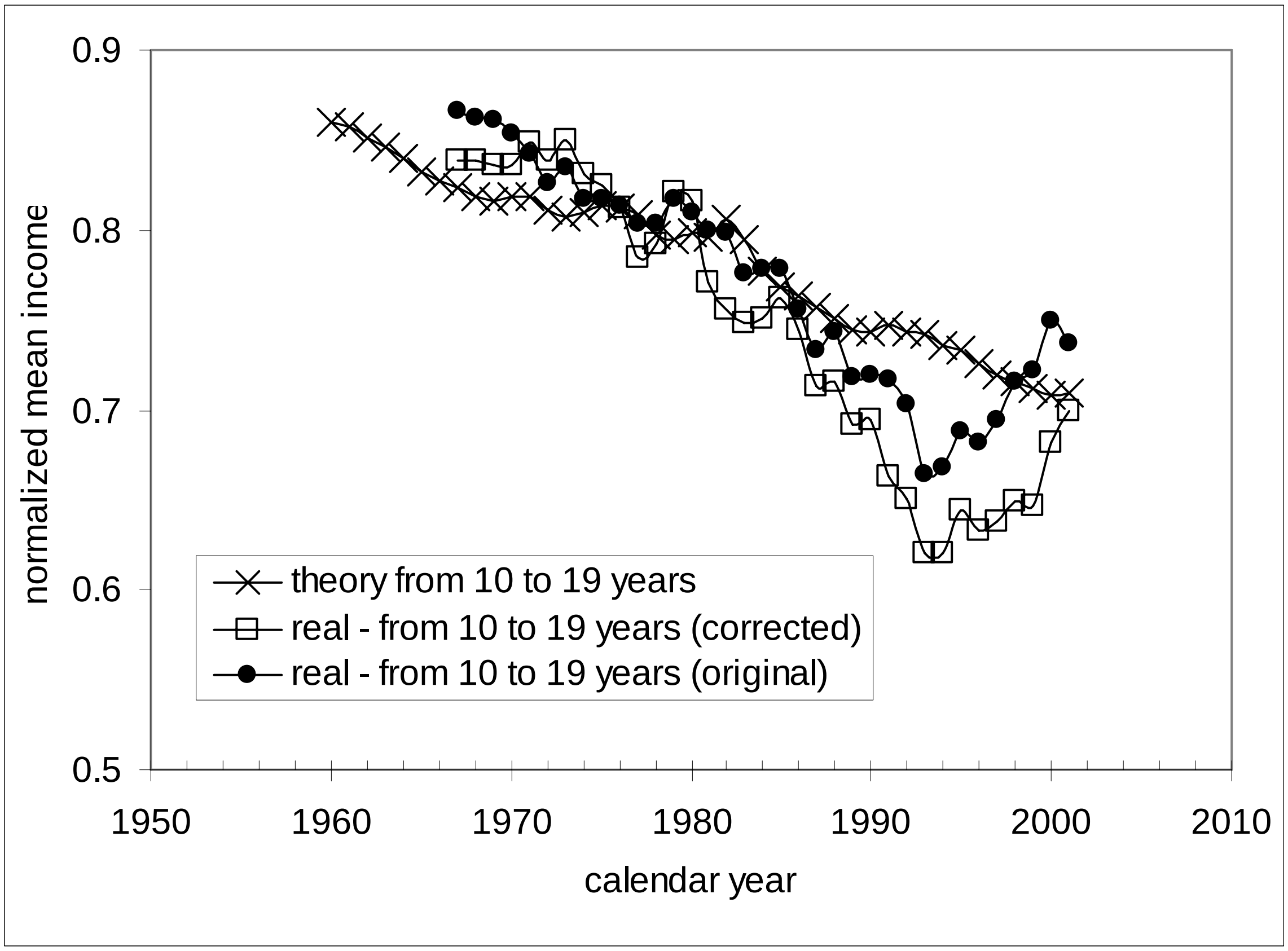

Fig. 20. Evolution of the observed and predicted (original and corrected for the population without income) average income value in the work experience group from 10 to 19 years, normalized to the peak average income over all the work experience groups.

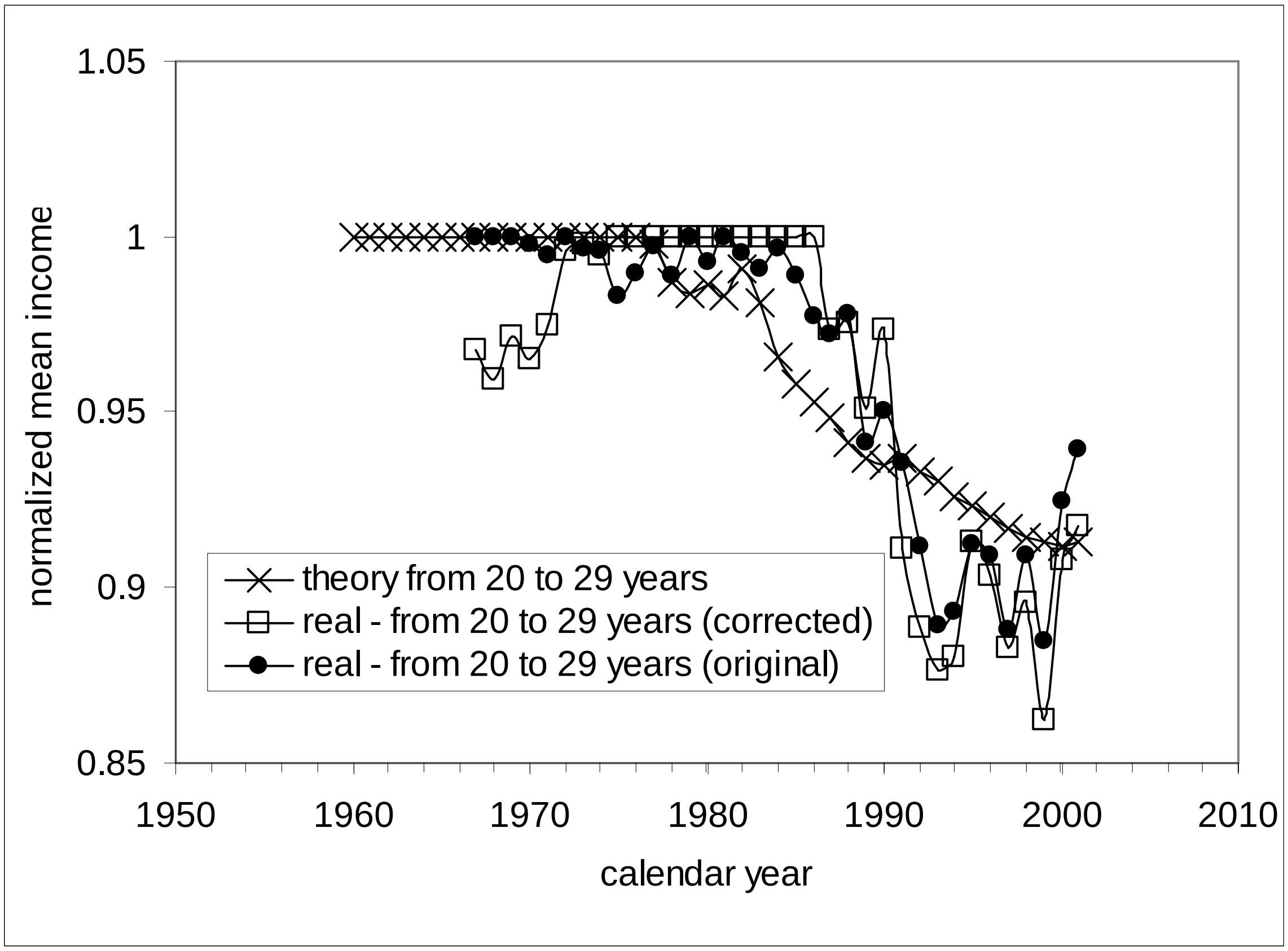

Fig. 21. Evolution of the observed and predicted (original and corrected for the population without income) average income value in the work experience group from 20 to 29 years, normalized to the peak average income over all the work experience groups.

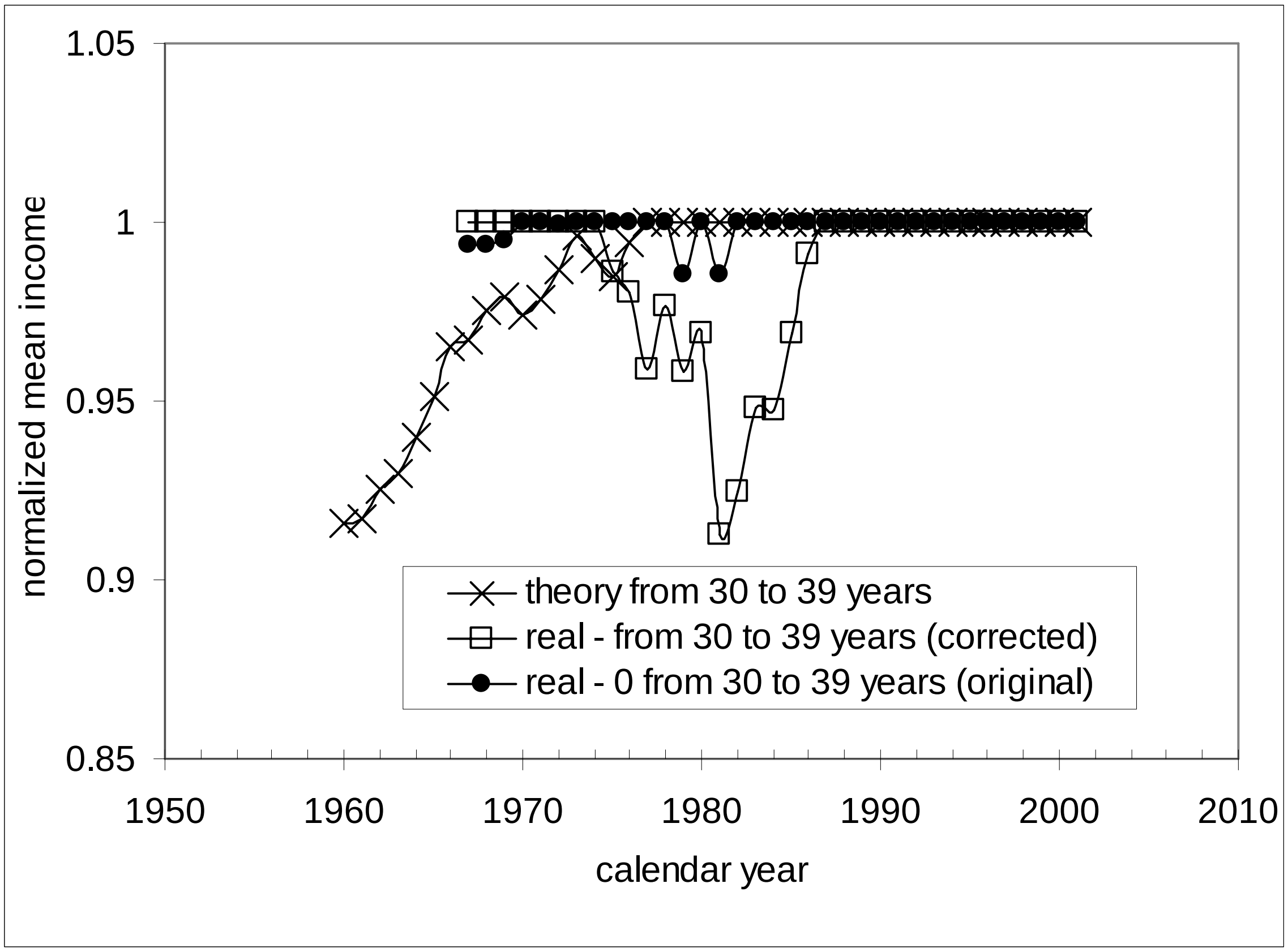

Fig. 22. Evolution of the observed and predicted (original and corrected for the population without income) average income value in the work experience group from 30 to 39 years, normalized to the peak average income over all the work experience groups.

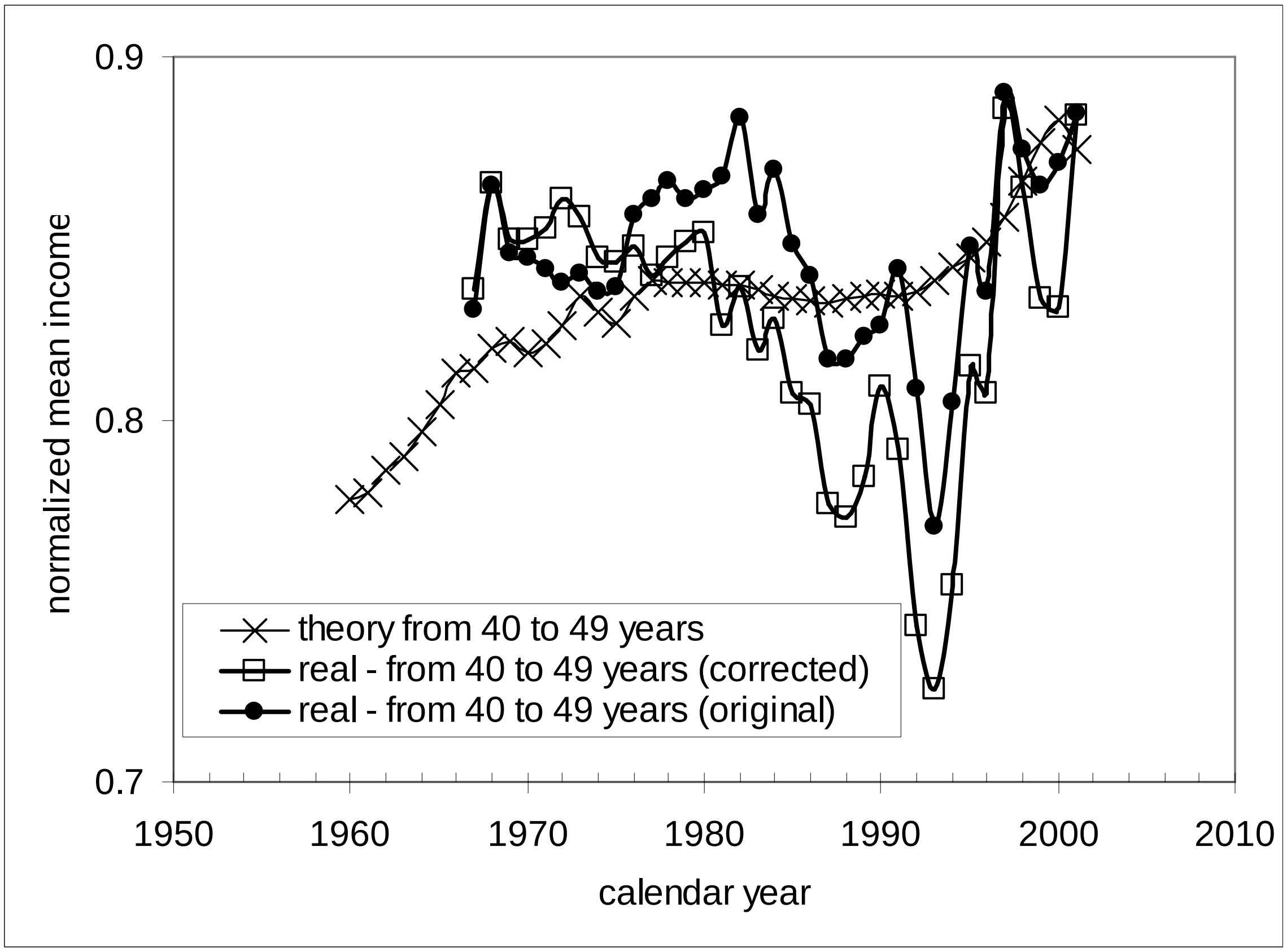

Fig. 23. Evolution of the observed and predicted (original and corrected for the population without income) average income value in the work experience group from 40 to 49 years, normalized to the peak average income over all the work experience groups.

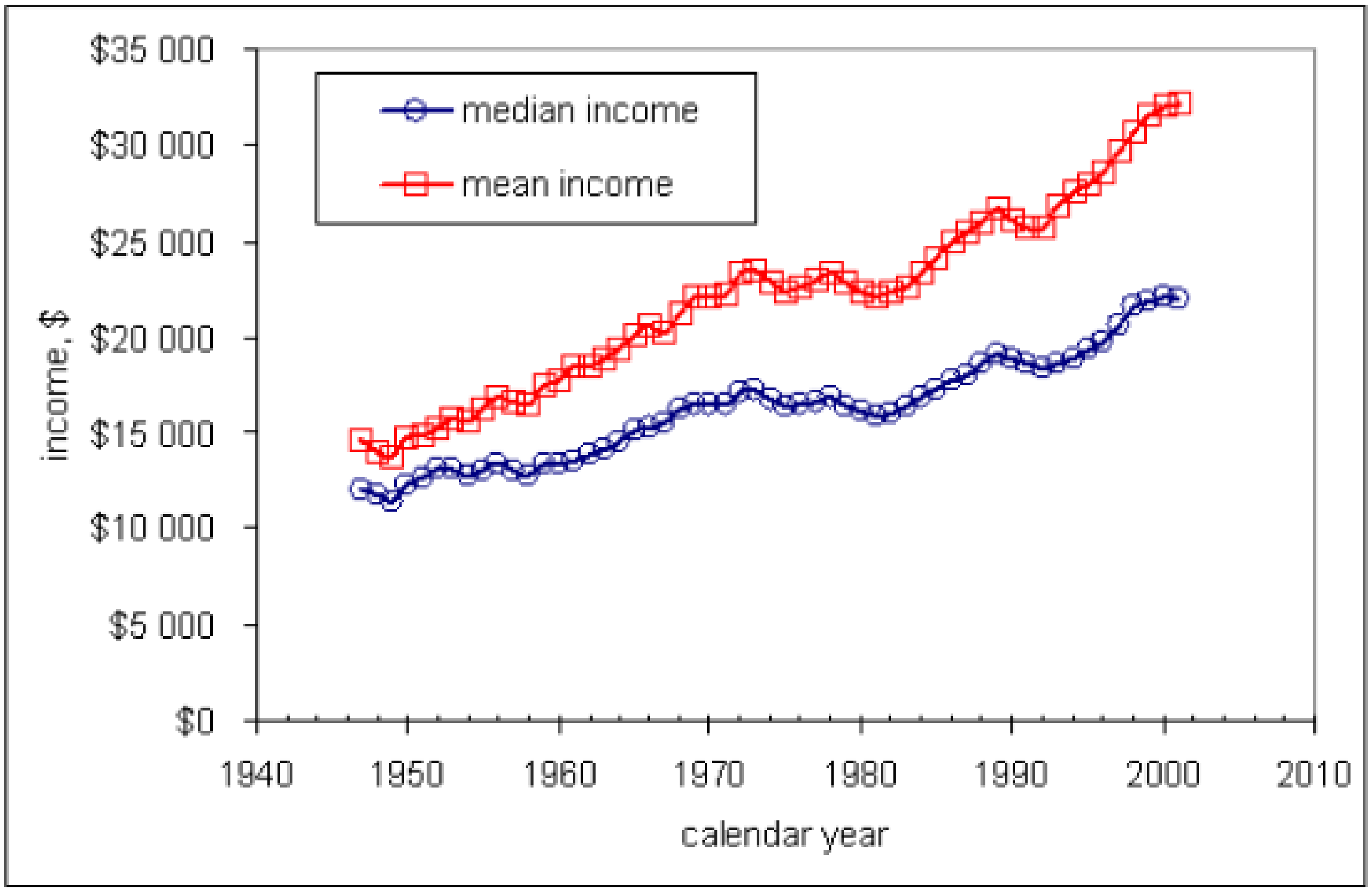

Fig. 24. Evolution of the observed mean and median income in the USA between 1947 and 2002.

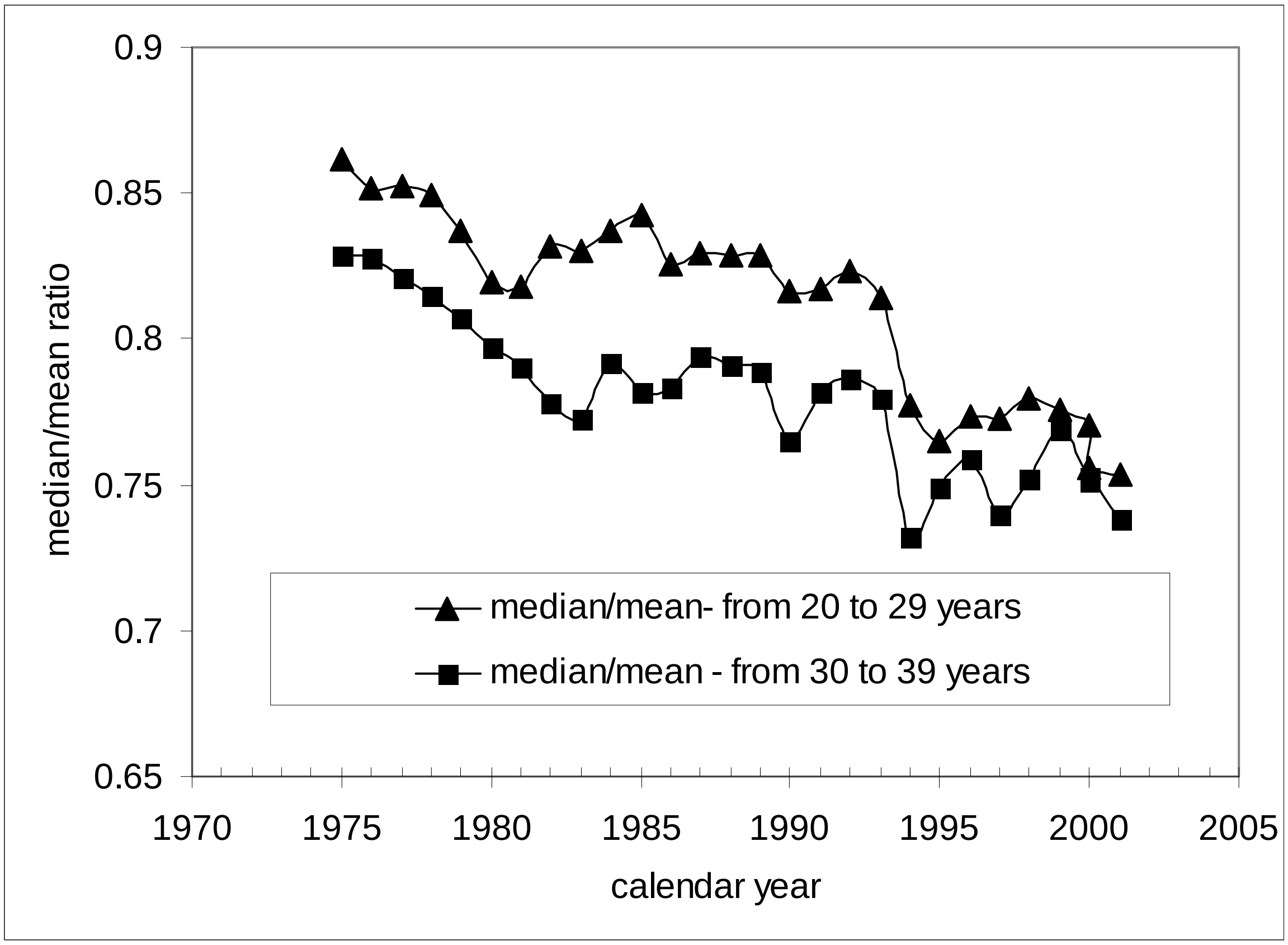

Fig. 25. Evolution of the median /mean income ratio in the work experience groups from 20 to 29 years and from 30 to 39 years. A decrease of the ratio in both groups is observed from 1974 to 2002.

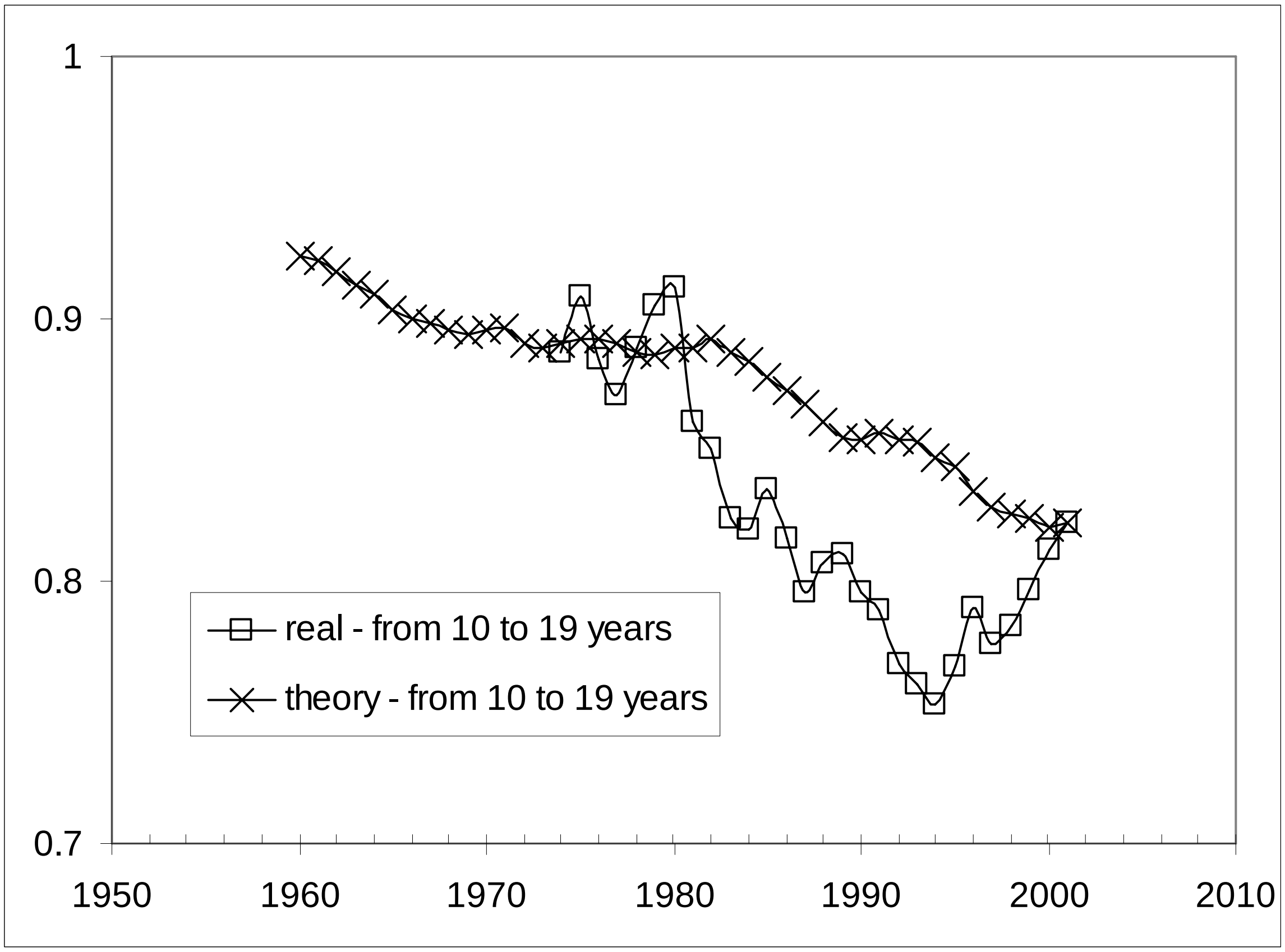

Fig. 26. Evolution of the observed and predicted median income value in the work experience group from 20 to 29 years, normalized to the peak average income over all the work experience groups.

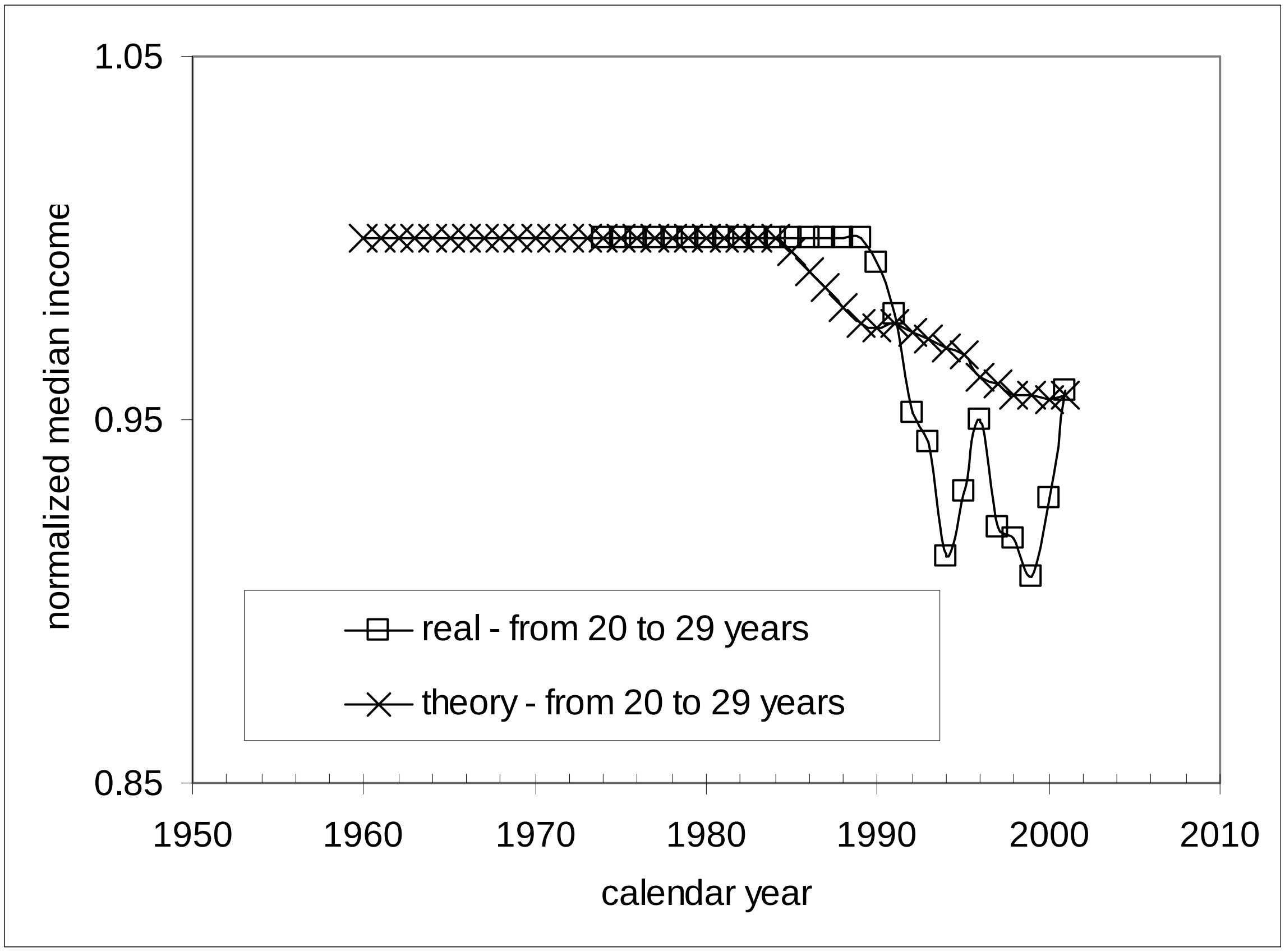

Fig. 27. Evolution of the observed and predicted median income value in the work experience group from 30 to 39 years, normalized to the peak average income over all the work experience groups.

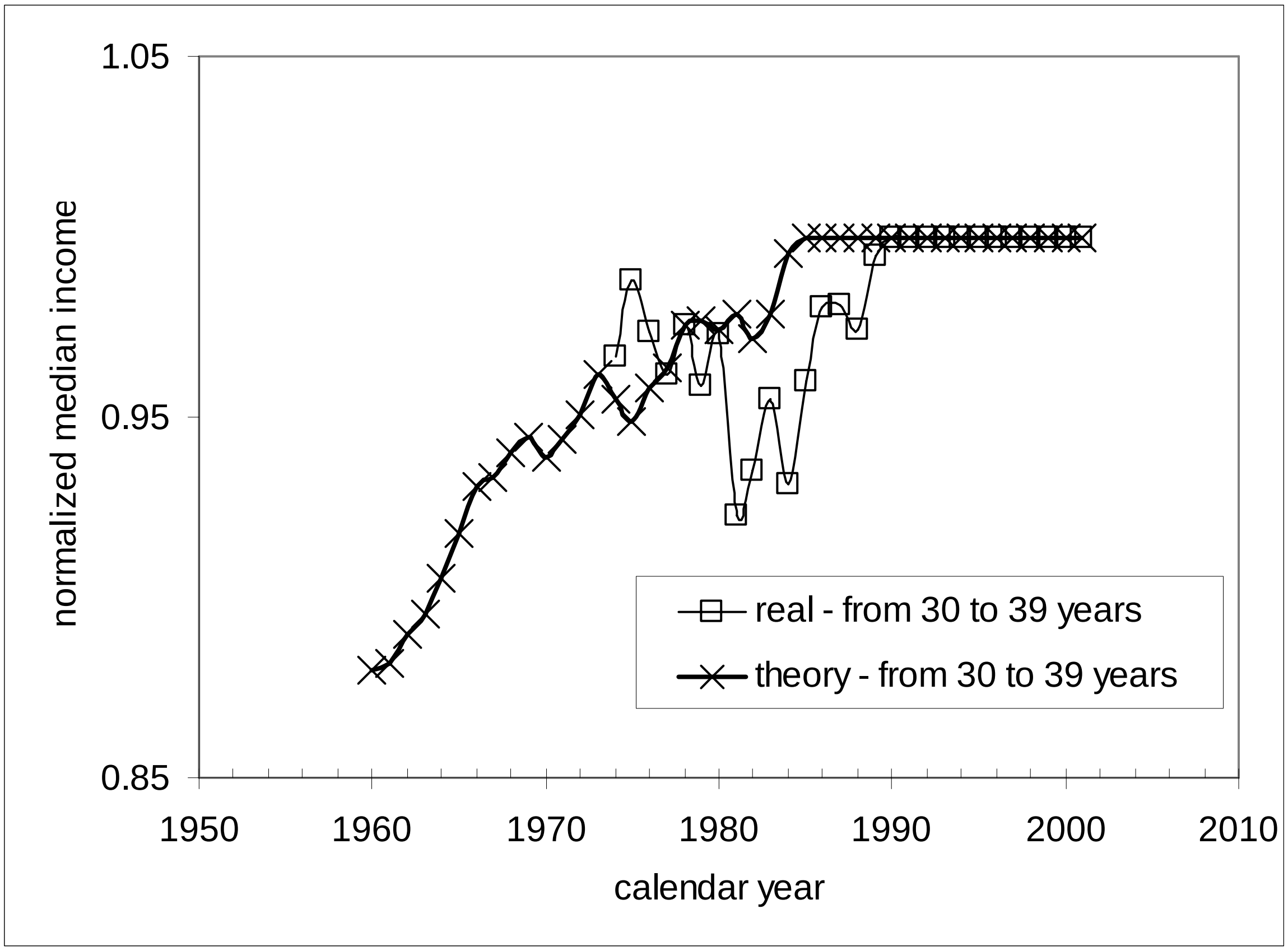

Fig. 28. Evolution of the observed and predicted median income value in the work experience group from 40 to 49 years, normalized to the peak average income over all the work experience groups.

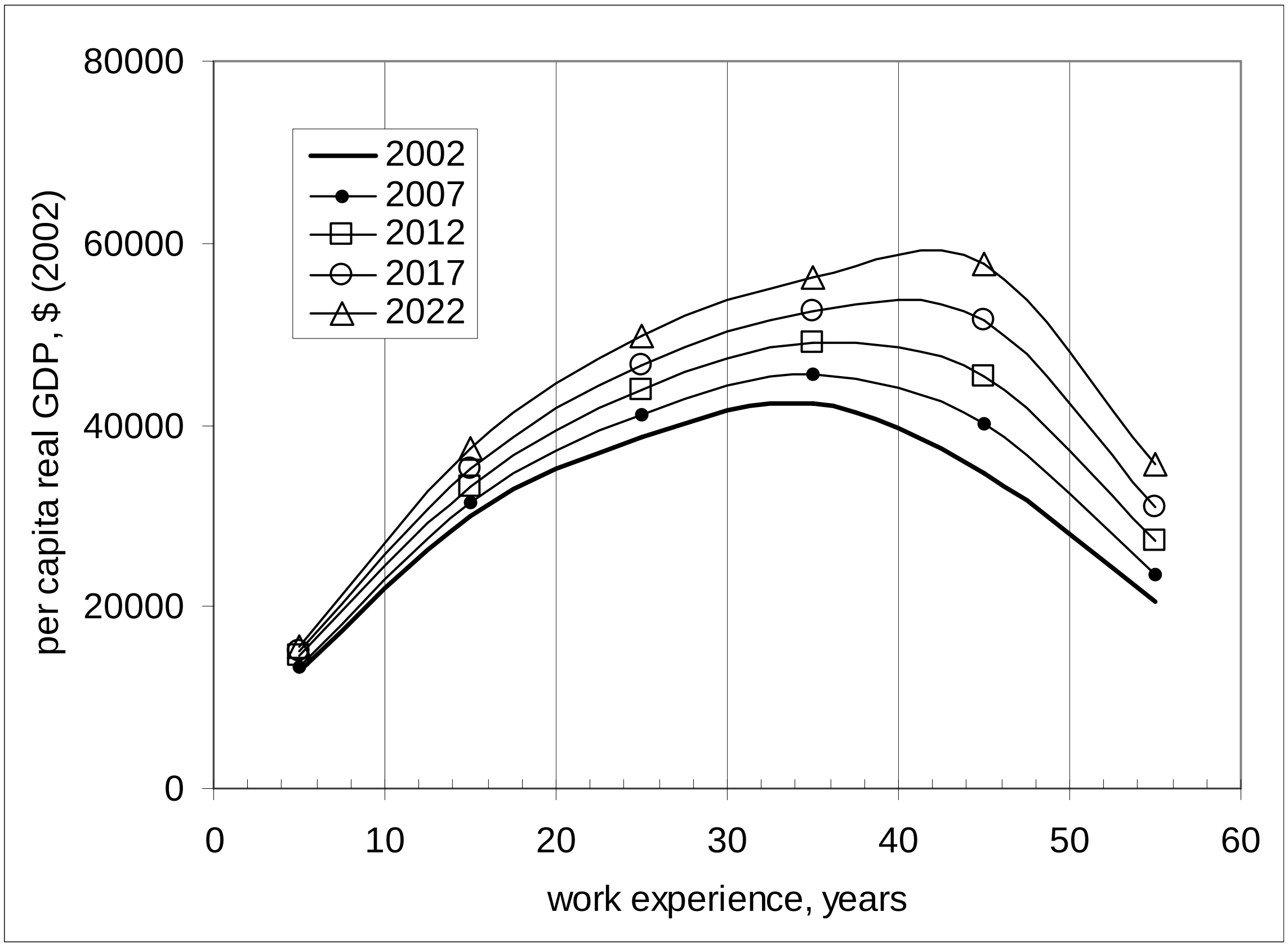

Fig. 29. Evolution of the mean income distribution. The growth rate of per capita real GDP is 1.6% per year. Population projections provided by the US Census Bureau are used.

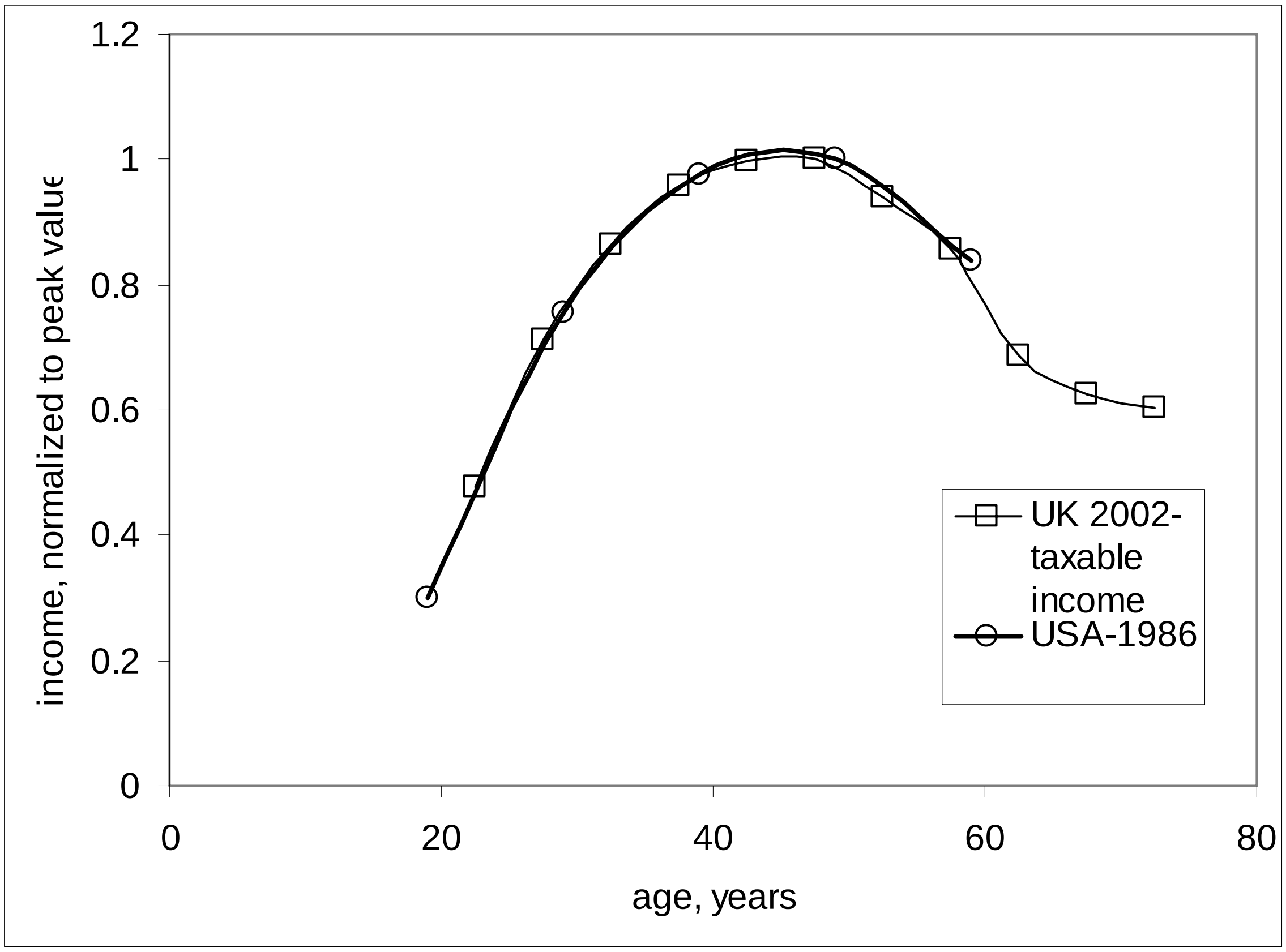

Fig. 30. Mean individual taxable income (Gross) in UK (current prices) in 2002 and the US average income distribution in 1986

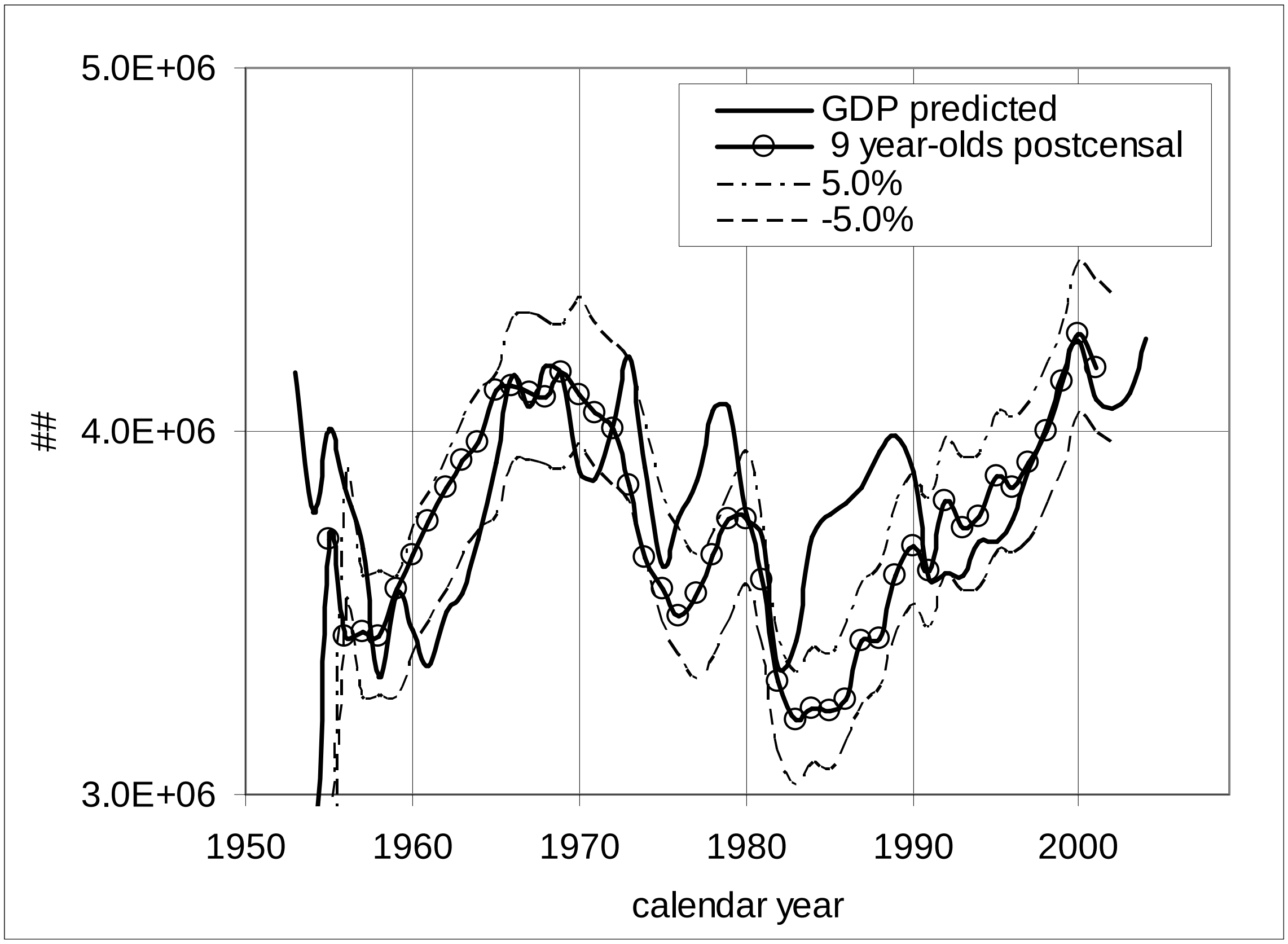

Fig. 31. Comparison of the US Census Bureau estimated number of 9-year-olds with that calculated from the real GDP per capita growth rate according to relationship (3). Five percent accuracy for the estimated values is assumed.